\documentstyle[11pt,aaspp4]{article}

\received{2 July 1996}
\accepted{26 October 1996}

\slugcomment{to appear in ApJ}

\begin{document}

\title{Evolution of Structure in the Intergalactic Medium\\
		and the Nature of the Ly$\alpha$ Forest}

\author{HongGuang Bi and Arthur F. Davidsen}
\affil{Center for Astrophysical Sciences, Department of Physics \& Astronomy\\ 
The Johns Hopkins University, Baltimore, MD 21218\\
received: 2 July 1996; accepted: 26 October 1996}

\begin{abstract}
We have performed a detailed statistical study of the evolution of structure 
in a photoionized intergalactic medium (IGM) using analytical simulations to 
extend the calculation into the mildly non-linear density regime found to 
prevail at $z = 3$. Our work is based on a simple fundamental conjecture: 
that the probability distribution function of the density of 
baryonic diffuse matter 
in the universe is described by a lognormal (LN) random field. The LN 
distribution has several attractive features and follows plausibly from the 
assumption of initial linear Gaussian density and velocity fluctuations at 
arbitrarily early times. Starting with a suitably normalized
power spectrum of primordial fluctuations in a universe dominated by
cold dark matter (CDM), we compute the behavior of 
the baryonic matter, which moves slowly toward minima in the dark matter 
potential on scales larger than the Jeans length. We have computed two models
that succeed in matching observations. One is a non-standard CDM model 
with $\Omega =1$, $h=0.5$ and $\Gamma = 0.3$, and the other is a low
density flat model with a cosmological constant (LCDM), with $\Omega = 0.4$,
$\Omega _{\Lambda} = 0.6$ and $h=0.65$. In both models,
the variance of the density 
distribution function grows with time, reaching unity at about $z = 4$,
where the simulation yields spectra that closely resemble the Ly$\alpha$
forest absorption seen in the spectra of high $z$ quasars. The calculations 
also successfully predict the observed properties of the Ly$\alpha$ forest
clouds and their evolution from $z = 4$ down to at least $z = 2$, assuming a 
constant intensity for the metagalactic UV background over this redshift 
range. {\em 
However, in our model the forest is not due to discrete clouds, but 
rather to fluctuations in a continuous intergalactic medium.}  At $z = 3$, 
typical clouds with measured neutral hydrogen column densities $N_{HI} = 
10^{15.3}, 10^{13.5}$, and $10^{11.5}$ cm$^{-2}$ correspond to fluctuations 
with mean
total densities approximately 10, 1, and 0.1 times the universal mean baryon 
density. Perhaps surprisingly, fluctuations whose amplitudes are less than 
or equal to the mean density still appear as ``clouds" because in our model 
more than 70\% of the volume of the IGM at $z = 3$ is filled with gas at 
densities below the mean value.
 
We find that the column density distribution of Ly$\alpha$ forest lines can 
be fit to $f(N_{HI}) \propto N_{HI} ^{-\beta}$, with $\beta = 1.46$ in 
the range $12.5 < \log N_{HI} < 14.5$, matching recent Keck results. At 
somewhat higher column densities the distribution steepens, giving $\beta = 
1.80$ over the range $14.0 < \log N_{HI} < 15.5$, and matching earlier 
observations for these stronger lines. The normalization of the line numbers 
in our model also agrees with observations if the total baryon density is 
$\Omega_b = 0.015 h^{-2}$ and the ionizing background intensity is $J_{21} =
0.18$. Alternatively, if $J_{21} = 0.5$ as recently estimated for the 
background due to observed quasars at $z = 2.5$, then $\Omega_b = 0.025 
h^{-2}$ yields the observed number of Ly$\alpha$ lines and the observed mean
opacity. The model predicts 
that about 80\% of the baryons in the universe are associated with Ly$\alpha$ 
forest features with $13 < \log N_{HI} < 15$ at $z = 3$, while 10\% are in 
more diffuse gas with smaller column densities and 10\% are in higher column 
density clouds and in collapsed structures, such as galaxies and quasars. Our 
model requires that absorbers at $z = 3$ with column densities higher than 
about $10^{16}$ cm$^{-2}$ -- Lyman limit systems and damped Ly$\alpha$ 
systems -- represent a separate population that has collapsed out of the IGM.
 
We find the number density of forest lines is $dN/dz = 75 [(1+z)/4] ^ {2.5}$
for lines with $EW > 0.32\AA$, also in good agreement with observations. We 
fit Voigt profiles to our simulated lines and find a distribution of $b$ 
parameters that matches that obtained from similar fits to real spectra. The 
effective opacity in the Ly$\alpha$ forest is found 
from the model to be $\tau_{eff} = 0.26 [(1+z)/4] ^ {3.1}$, again in good 
agreement with observations. The exponents for the evolution of the number of 
lines and the effective opacity do not simply differ by 1.0, as in the 
standard cloud picture, owing to saturation effects for the stronger lines. 
This also explains why the effective opacity evolves more slowly
than $(1+z) ^{4.5}$, which is expected for the Gunn-Peterson effect in a 
uniform medium with $J =$ constant (and $\Omega = 1$).

We compare the spectral filling factor for our CDM and LCDM models with
observations of the Ly$\alpha$ forest in HS 1700+64 and find good agreement.
Similar calculations for a standard CDM model and a cold plus hot dark matter
model fail to match the observed spectral filling factor function.
We have also compared a number of predictions of our analytical model with
results of numerical hydrodynamical calculations (Miralda-Escud\'e et al.
1996) and again find them to be in good agreement. The lognormal hypothesis,
coupled with otherwise attractive CDM-dominated cosmological models,
appears to provide a plausible and useful description of the distribution
of photoionized intergalactic gas and provides a new estimate of the
baryonic density of the universe.
\end{abstract}

\keywords{intergalactic medium, quasars: absorption lines,
cosmology: large-scale structure of the Universe, hydrodynamics, dark matter}

\section{Introduction}

Ever since the discovery of quasars with redshifts sufficiently large to
bring the Ly$\alpha$ resonance line of neutral hydrogen into the optical
band, the possibility of using this line to probe conditions in the 
intergalactic medium (IGM) has engendered intense interest. Gunn \& Peterson
(1965) showed that a diffuse, uniform IGM having the critical density must
have been very highly ionized at $z=2$ to avoid producing a very large HI
opacity at wavelengths just below that of the quasar's Ly$\alpha$ emission
line. Motivated by the Gunn-Peterson idea, Bahcall \& Salpeter (1965) pointed
out that the Gunn-Peterson effect could be used to probe the IGM at various
redshifts (and for other elements), and also suggested that gas that was not 
uniformly distributed would produce discrete Ly$\alpha$ absorption lines.
They were then thinking specifically of gas clumped into clusters of galaxies,
but a small generalization will allow us to refer to any localized Ly$\alpha$
absorption in the IGM as the ``Bahcall-Salpeter effect''. The subsequent
discovery of large numbers of discrete absorption lines at wavelengths below
redshifted Ly$\alpha$ (Lynds 1971) lead to the description of this
phenomenon as the ``Ly$\alpha$ forest''. Their large numbers
showed that these forest lines could not be associated with galaxy clusters,
and after a period of uncertainty as to whether the lines might be associated
with local material ejected from the quasars at high velocities, the idea
took hold that the Ly$\alpha$ lines arise in discrete intergalactic 
clouds at various cosmological redshifts along the line of sight 
(Sargent et al. 1980; Weymann, Carswell \& Smith 1981; Blades,
Turnshek \& Norman 1988). Observations at relatively high spectral resolution
revealed no continuous absorption between the discrete lines, placing 
strong limits on the Gunn-Peterson effect, and leading to the idea that
the Ly$\alpha$ ``clouds'' might be confined by the pressure of a diffuse, 
very hot, intercloud medium. Over time this two
phase notion lead to the identification of the {\em intercloud}
medium with the {\em intergalactic} medium, while the Ly$\alpha$ clouds
became a somewhat separate subject of study.

In this paper we wish to resurrect the concept of a unified, though not 
uniform, intergalactic medium, consisting of all the gas that has not
collapsed into systems with densities much more than an order of magnitude
above the mean baryon density of the universe. In the model we have 
developed, relatively small amplitude fluctuations in the density of the
IGM can explain numerous aspects previously
attributed to the discrete cloud picture. We suggest
that the Ly$\alpha$ forest is nothing more than the observable manifestation
of the Bahcall-Salpeter effect, which is in turn just a generalization
of the Gunn-Peterson effect to the case of a nonuniform intergalactic
medium.

Our approach has the merit of great simplicity, with numerous predictions 
following from a small number of fundamental assumptions. The most important
of these assumptions
is that the probability distribution function of the density of
diffuse matter in the universe is described by a lognormal (LN) random
field. The LN distribution has several attractive features and follows 
plausibly from the assumption of initial linear Gaussian density and
velocity fluctuations at arbitrarily early time (Coles \& Jones 1991).
It has the correct asymptotic behavior on large scales or at very early 
times where the IGM evolves linearly, and it matches the  
isothermal hydrostatic solution on very small scales such as that
for intracluster gas. We also find it to be roughly correct
when compared to the statistics of the density distribution
in some recent numerical simulations (see \S 2.1).

We further assume that the IGM is photoionized by a metagalactic UV radiation
field whose intensity is sufficient at $z=2 - 4$ to reduce the neutral
fraction of hydrogen to less than $10^{-5}$ for gas at the mean density.
With a suitable choice of $\Omega _b$, taken to be the nominal baryon
density from standard Big Bang nucleosynthesis, for example, this assumption
can be satisfied using the intensity provided by the known quasar population.

Our investigation of the model developed here has been motivated by the
recent detection of the HeII Gunn-Peterson/Bahcall-Salpeter effect 
(Miralda-Escud\'e \& Ostriker 1990: Davidsen 1993; and Miralda-Escud\'e 1993)
with the Hubble Space Telescope (Jakobsen et al. 1994; Tytler et al. 1995) 
and the Hopkins Ultraviolet Telescope (Davidsen \& Fountain 1985;
Davidsen 1993; Davidsen, Kriss \& Zheng 1996,
hereafter DKZ). The HUT observation of HS1700+64 yielded an effective
HeII Ly$\alpha$ opacity $\tau _{HeII} = 1.0$ at $z=2.4$. This opacity is
too large to be accounted for using the canonical Ly$\alpha$ 
cloud picture, and lower column density clouds ($N_{HI} \le 10^{13}$
cm$^{-2}$) and/or diffuse gas are needed to explain more than half of the
measured He II opacity toward HS1700+64 (DKZ; Zheng, Davidsen \& Kriss
1996). The very low density gas in our model contributes very little
to the H I opacity, but because the number ratio between
He II ions and H I atoms 
resulting from
photoionization by the UV background is large (i.e. $n(HeII)/n(HI) \simeq
100$ -- 200), it
contributes very significantly to the He II opacity. In this paper we 
develop our IGM model and concentrate on comparisons with observations
of the HI Ly$\alpha$ forest and with other theoretical work.
Application of the model to the HeII observations will be
discussed in Davidsen
et al. (1996). Some of the ideas employed in our model have
previously been discussed by Bi, B\"orner \& Chu (1992), Bi (1993)
and references therein.

Ly$\alpha$ clouds were suggested to be confined either by an intercloud medium
(pressure confinement, e.g. Sargent et al. 1980), or by dark matter mini-halos
(Rees 1986). A very hot baryonic medium would distort the blackbody
spectrum of the cosmic microwave background radiation through the
Sunyaev-Zel'dovich effect. The recent upper
limit of the so-called y-parameter has ruled out any cosmologically
distributed component of temperature greater than $10^6 K$
(Wright et al. 1994). Thus, the pressure confinement seems unfavored by 
the COBE results.  In the original mini-halo
model, baryons are presumed to reside statically in dark 
matter potential wells.
In our model, we allow baryons to move in or out of the
wells, depending on the strength of the well. 
So the dark matter potential
wells are more like magnets that direct the motion of
baryons but do not trap them. 
Removing the confinement requirement will introduce a larger
cloud size as observed by Dinshaw et al. (1994) and Bechtold et al. (1994).

Historically, there have been several ideas related to unconfined clouds.
Black (1981) and Oort (1981) suggested that the 
Ly$\alpha$ forest might represent relic primordial fluctuations having a 
characteristic length scale $0.6 - 1.4$ Mpc. By assuming UV photoionization
heating, Ostriker \& Ikeuchi (1983) calculated the temperature of the IGM
to be about $1 - 3 \times 10^4$ K, very close to that inferred from the widths
of the Ly$\alpha$ lines. Even in the mini halo model,
the clouds may act like transient phenomena
because the halos merge frequently at high redshift. 

The first non-static cloud model was studied by Bond, Szalay
\& Silk (1988) who suggested that clouds were in a free expansion phase. 
McGill (1991) illustrated preliminarily that an IGM with the 
fluctuation variance $\sigma ^2 = 1$ could also produce line-like 
absorptions in 
quasar spectra through caustic structures in its peculiar velocity field. 
Bi, B\"orner \& Chu (1992) and Bi (1993) considered a more physical IGM 
evolution in the hierarchical clustering model and the cold-dark-matter
model. They found that such a median fluctuated IGM contains the 
correct number of density clumps, which have sizes comparable to
the Jeans length, and would produce Ly$\alpha$ absorption lines compatible 
with what are observed. The velocity caustics are a secondary effect 
when compared to the density fluctuation. The similarity between two nearby 
absorption profiles, such as seen in 
gravitationally-lensed quasar spectra, can be 
consistently accounted for by coherent clustering in the density too. 
Therefore
they suggested that such an alternative phase of the IGM should have played 
a role in the Ly$\alpha$ forest as important as the conventional clouds.
The reason that the $\sigma = 1$ IGM can produce distinguishable line
features is that the absorption profile is an exponential function of the 
optical depth $\tau$, which in turn is approximately proportional to the
square of the 
baryonic density. Therefore a small enhancement (i.e. an overdensity
over the local background) in the 
IGM would result in a large change in $\exp (-\tau )$ at $\tau \simeq 1$
and then appear as a deep absorption line.

The actual fluctuation picture can be obtained by cosmological N-body and 
hydrodynamical simulations. Recent simulations by Cen et al. (1994), Zhang, 
Anninos \& Norman (1995), Hernquist et al. (1996), and Miralda-Escud\'e et al 
(1996) have solved hydrodynamical equations from 
first principles and have set up a firm evolutionary picture of the IGM. 
Although they use different techniques
and cosmological models, all the simulations 
indicate a median fluctuated IGM instead of discrete clouds.
For instance, from an analysis of published simulations
we found the variance of the IGM density is just 
about 1 at $z=3$ in Cen et al. (1994) and Miralda-Escud\'e et al (1996).
The density, temperature and thermal pressure of the medium are continuous 
fields that cannot be attributed simply to
gravitational confinement or pressure confinement. Yet the Ly$\alpha$ forest
can be produced naturally. 

What are the geometry, ionization degree, thermal state and Ly$\alpha$
absorption properties of absorbers in the IGM ? How do the statistical 
quantities evolve in redshift ? The present paper will study these 
questions in detail. Related work can be found in Miralda-Escud\'e
et al. (1996, hereafter MCOR). 
Because of our analytical approach, our simulation can be performed in
1-D, i.e. along one line of sight. This enables us to have a simulation 
box size 
almost arbitrarily large,  which is necessary for large-scale fluctuation
components, as well as to have a very small grid in the same 
simulation for revealing fine structures {\em
within} Ly$\alpha$ clouds. Different cosmological models with different
choices of free parameters can be computed relatively quickly
to check various diagnostics in our IGM simulation.

Flat universe models dominated by cold dark matter, 
with or without a cosmological 
constant, will be considered in this paper. We will discuss the assumption 
of the lognormal field, and our simulation procedure in Section 2. 
In Section 3.1, the evolution of the volume filling factor and 
cumulative mass fraction for different density thresholds will show why 
the IGM should be regarded as a diffuse and continuous medium. Several 
statistical properties of Ly$\alpha$ absorbers are discussed in Section 3.2, 
including 
characteristic size, mean density and neutral fraction. Velocity and 
density structures within clouds will be studied in Section 3.3. 
The evolution of HI opacity, cloud number counts
$dN/dz$, and the HI column density distribution will be discussed 
and compared to observations in Section 4. 
In Section 5, we compare simulation samples with a high
resolution Keck observation of
HS1700+64. They agree surprisingly well with each other, not only 
in general appearance
but also in their $b$ parameters, spectral filling factors
and spectral autocorrelation coefficients. We have also made some
comparisons to MCOR's simulation in that section. Section 6 summarizes
our conclusions.  

\section{The Lognormal Density Field : Theories and Structures}
 
\subsection{Constructing the IGM model}

The density distribution of the IGM is assumed to be a lognormal 
random field,
\begin{equation}
n (\vec{x}) = n_0 \exp[ \delta _0 (\vec{x}) - \frac{<\delta_0 ^2>}{2} ],
\end{equation}
where $n_0$ is the mean number density and $\delta _0$ is a Gaussian 
random field derived from the density contrast $\delta _{DM}$ of
dark matter : 
\begin{equation}
\delta _0(\vec{x}) \equiv \frac{1}{4\pi x_b ^2} \int 
\frac{\delta _{DM} (\vec{x}_1)}
{|\vec{x} - \vec{x}_1|}e^{-\frac{|\vec{x} - \vec{x}_1|}{x_b}} d\vec{x}_1, 
{\mbox{\ \ \ or  }}
\delta_0 (\vec{k}) \equiv \frac{\delta _{DM} (\vec{k})}{1+x_b^2k^2},
\end{equation}
in the comoving space or the Fourier space,
respectively. The comoving scale $x_b$ is defined by
\begin{equation}
x_b \equiv \frac{1}{H_0} (\frac{2\gamma k T_m }
{3\mu m_p \Omega (1+z)}) ^{\frac{1}{2}},
\end{equation}
where $T_m$ and $\mu$ are the mean temperature and molecular weight of the 
IGM; $\Omega$ is the cosmological density parameter of total mass; and
$\gamma$ is the ratio of specific heats. The scale $x_b$ is $2\pi$
times smaller than the standard Jeans length $\lambda _b \equiv
v_s (\pi /G \rho _c )^{1/2}$, 
where $v_s$ is the sound speed in the IGM and $\rho _c$ is the total
cosmological mass density (Peebles 1980).

The LN approach was introduced by Coles \& Jones (1991) to describe
the non-linear evolution of dark matter. Here we apply it to the baryonic
matter in the IGM.
We first discuss the physical background of the LN approach.

At early epochs $z \gg 1$ 
or on large scales $x\to \infty$, fluctuations are small, so we have
$\delta (\equiv n/n_0 -1) \simeq \delta _0 $ in Eq. 1. This is just the 
expected linear evolution of the IGM (Peebles 1974; Fang et al. 1993).
On very small scales, we have $|\vec{x} - \vec{x}_1| \ll x_b$ in Eq. 2; 
thus Eq. 1 becomes the well-known isothermal hydrostatic solution, which
describes highly clumped structures such as intracluster gas,  
\begin{equation}
n \propto \exp (-\frac{\mu m_p}{\gamma k T} \psi _{DM}),
\end{equation}
where $\psi _{DM}$ is the dark matter potential (Sarazin \& Bahcall 1977).

During the linear evolution of dark matter fluctuations,
$\psi _{DM}$ is constant, being as small as that at the decoupling time.
Therefore, one expects that the potential would remain linear much
longer than the density; 
consequently one can simply use the unchanged primordial
potential in the whole calculation. This frozen potential approximation 
has been tested and found to work
well against N-body simulations, and is apparently superior to the Zel'dovich
approximation or the frozen-flow approximation (Brainerd, Scherrer \&
Villumsen 1983; Bagla \& Padmanabhan 1994).
As long as the dark matter potential is fixed, baryons will move
toward the minima in the potential, which are 2-dimensional pancakes.
If the evolution is long enough, the final configuration will be
the isothermal hydrostatic solution.

In the LN distribution, the 3-D configuration of the IGM is 
uniquely determined by the primordial distribution. This can be compared
to a recent non-linear clustering model, the peak patch theory
(Bond, Kofman \& Pogosyan 1996). N-body simulations indicated that
the final non-linear density field can be fairly reconstructed from
individual ``peak patches" existing in the original distribution.
The LN approach can be viewed alternatively as a very simplified 
representation of the theory, in which all small scale collisons 
between nearby particles are negelected.

Among all smooth functions that can link the known linear
solution and the hydrostatic solution asymptotically, 
the LN function is the simplest one. Furthermore, it
always defines a positive IGM density. In contrast, any 
smooth polynomial function of $\delta _0$ will output negative density $n$ 
when $\delta _0 \to -\infty$. Because $\delta _0$ is identified as the 
density contrast in the linear evolution, Eq. 1 transforms the extrapolated 
linear density to the nonlinear counterpart. 

Our conjecture can be tested by 
comparisons with hydrodynamical simulations. We have made
a statistical histogram of $\log \frac{n}{n_0}$ at $z=3$ in MCOR's Fig. 7, 
which turns out to be fitted well by a Gaussian probability function with
variance $\sigma ^2 = 1.0$ and the mean -0.52.  Note that according to 
Eq. 1, the mean should be equal to $-\frac{1}{2} \sigma ^2$. Based on the 
same cosmological model, our analytical calculation leads to $\sigma = 1.25$,
which is only slighter greater. We found that the results discussed later 
in the paper are not altered substantially if the smaller variance is 
adopted. Hence, we suggest that the LN model is a simple, yet acceptable 
model for the density distribution of the IGM. 

A Gaussian random field is uniquely determined by its power spectrum.
We will use the cold-dark-matter power spectrum in our simulation of 
$\delta _0$. This is (Efstathiou, Bond \& White 1992)
\begin{equation}
P(k) \propto k/(1+(6.4 \frac{k}{\Gamma} + (3.0 \frac{k}{\Gamma})^{1.5}
                   + (1.7 \frac{k}{\Gamma})^2)^{1.13}) ^{\frac{2}{1.13}} ,
\end{equation}
where $\Gamma = \Omega h$, and $h$ is the normalized Hubble parameter. The 
amplitude of the spectrum can be normalized by the COBE rms quadruple 
$Q_{rms} = 18\mu $K. If normalized in this way with $\Gamma = 0.5$, the 
model $\Omega = 1$ and $h=0.5$ predicts a correlation function higher than 
that observed on scales of $\sim 10 h^{-1}$Mpc (Efstathiou at al. 1992). 
Therefore, Efstathiou at al. suggested replacing $\Gamma = 0.5$ with 
$\Gamma = 0.3$ in Eq. 5. We will call this {\em the CDM model} in this 
paper. Another possibility to get a lower $\Gamma$ is to introduce a 
cosmological constant $\Lambda$ so that $\Omega + \Omega _{\Lambda} = 1$. 
MCOR's simulation is based on such a model. We will use the same parameters 
and normalization as MCOR's, i.e. $\Omega = 0.4$, $\Omega_{\Lambda} = 0.6$, 
$h = 0.65$ and the present fluctuation variance
0.79 on the scale of 8$h^{-1}$Mpc. We call this {\em the LCDM model}. 
We note that if the LCDM model is normalized by the COBE value, the
fluctuation variance will be increased by approximately 30\% and its
fit to the observations will not be as good. All our simulations are 
performed equally for both the models. 

The mean IGM temperature, which is $1 - 3 \times 10^4$ K between $z=5$ and 
$z=0$, can be calculated unambiguously when UV photo-ionization is the only 
heating source (Ostriker \& Ikeuchi 1983). The thermal equation of state
of the IGM is assumed to be polytropic, $T \propto n ^{\gamma -1}$. 
In the linear evolution, one can prove that $\gamma = \frac{5}{3}$; in 
MCOR's simulation (their Fig. 7), we find this relation is best fit by 
$\gamma \simeq \frac{4}{3}$, which is also preferred in the calculation
of Miralda-Escud\'e \& Rees (1993). In addition, the lowest temperature in 
MCOR does not drop to zero, but is limited at the photo-ionization temperature 
$\sim 10^{4}$ K. We will use $\gamma = \frac{4}{3}$, the constant temperature
$T_0 = 2\times 10^4$ K at the mean density $n_0$, and a minimum temperature
$10^4$ K to approximate the variation of $T$ with $n$ in the IGM. Our final 
results are very weakly dependent on the choice of $\gamma$ or the minimum
temperature. However, in the polytropic IGM, most mass resides in 
moderately high
density regions where the temperature is considerably higher than $T_0$. We 
should therefore use a density-averaged temperature $T_m$ in the calculation 
of the Jeans length to account for this effect. Because $\delta _0 (k) 
\propto T_m ^{-1}$ at large $k$ in Eq. 2, such a change is not trivial for 
the LCDM model, or at $z\le 1$ in both models, when the clustering is 
strong. This introduces a factor of 2 (or even greater) difference between 
$T_0$ and $T_m$. For instance, if $T_0$ is used in the calculation of the 
Jeans length, the resultant $\sigma$ of the IGM will be 
about 2.0 at $z=3$ in the LCDM model; this is too large when compared to 
MCOR's value of 1.0.  

Hydrogen atoms in the IGM are photo-ionized by the UV background radiation,
which is assumed to be 
$J(\nu ) = J_{21} \times 10 ^{-21} (\nu _0/\nu )^{1.5}$ ergs s$^{-1}$
Hz$^{-1}$ cm$^{-2}$ sr$^{-1}$, where $\nu _0$ is the frequency of the H I
ionization threshold. The
optical depth is proportional to the combination of $J_{21}$, $\Omega _b$
and $h$ through $\frac{(\Omega_b h)^2}{h J_{21}}$; therefore, we fix 
$\Omega _b$ to the primordial density, $\Omega _b = 0.015 h^{-2}$
(Walker et al. 1991) , and let
$J_{21}$ be a free parameter.  

The absorption optical depth at redshift $z_0$ is 
\begin{equation}
\tau (z_0) = \int ^{t_0} _{t_1} \sigma (\nu _{\alpha}\frac{1+z}{1+z_0})
             n_{HI} (t) dt, 
\end{equation}
\begin{equation}
\ \ \ \ \ \ = \frac{cI_a}{H_0} \int ^{y_1} _0 dy \cdot n_{HI} (y) \cdot \frac
{1}{\sqrt{\pi } (1+z) b} V(\alpha, \frac{z-z_0}{b(1+z_0)} + \frac{v(y)}{b});
\end{equation}
where $\sigma$ is the absorption cross section at the Ly$\alpha$ transition,
$y = x H_0/c$ is
the normalized comoving coordinate, $I_a = 4.45\times 10^{-18}$ 
cm$^2$, $V$ is the standard Voigt function, $b = \sqrt{2kT/mc^2}$ and $v(y)$ 
is
the peculiar velocity measured in units of $c$. There are two major effects 
of the velocity: 1) shifting lines to slightly different position and 2) 
altering line profiles by bulk motions through velocity gradients. The first 
effect is not important if we are not interested in clustering properties of 
lines. The second effect can make lines narrower, since in comoving space 
mass is falling toward the center of the cloud. Only clouds with $N_{HI} \le 
10^{11}$ cm$^{-2}$ are expanding in comoving space (see \S 3.3)
and consequently have 
broader line profiles, but these lines are not observable. In the present 
paper, we will use the linear estimate for the peculiar velocity. The success 
of the Zel'dovich approximation indicates that the velocity can be described 
by linear theory in the early non-linear stages. (Note the velocity is the 
first derivative of the potential and the density is the second derivative). 
 
\subsection{Simulation method}

Here is the summary of the simulation procedure. We first simulate 
$\delta _0(k)$ and $v (k)$ in a line of sight at discrete $k_i, 
i=1, 2, ..., N$ in the Fourier space. They are two correlated Gaussian
random fields and can be written as linear combinations of two 
independent Gaussian fields $w(k)$ and $u(k)$ of power spectra
$P_w(k)$ and $P_u(k)$, respectively (see Bi 1993 and Bi, Ge \& Fang 1995) :
\begin{equation}
\delta _0 (k, z) = D(z) (u(k) + w(k)),
\end{equation}
\begin{equation}
v(k, z) = F(z) \frac{H_0}{c} ik \alpha (k) w(k),
\end{equation}
where $D(z)$ and $F(z)$ are the linear growth factors for $\delta _0$ 
and $v$ at redshift $z$ (Peebles 1993). Functions $\alpha$, $P_w$ and 
$P_u$ are :
\begin{equation}
\alpha(k) = \frac{\int ^\infty _k P_{IGM}(q) q^{-3} dq}
            {\int ^\infty _k P_{IGM}(q) q^{-1} dq},
\end{equation}
\begin{equation}
P_w(k) = \alpha ^{-1} \int ^\infty _k P_{IGM}(q) 2\pi q^{-1} dq,
\end{equation}
\begin{equation}
P_u(k) = \int ^\infty _k P_{IGM}(q) 2\pi q dq - P_w(k),
\end{equation}
with $P_{IGM}(k) = P(k) / (1+x_b ^2 k^2)^2$ 
being the present 3-D power spectrum 
of the IGM. The corresponding distributions in the real line-of-sight
space can be
obtained by using a Fast Fourier Transform. 
Typically, $N = 2^{14} = 16384$ and the 
length of the simulation is $\Delta z = 0.2$
in redshift, corresponding to the grid size 4.5 
$h^{-1}$kpc and the simulation box size 75 $h^{-1}$Mpc
in comoving space at $z=3$ in the  $\Omega =1$ model.
The grid is fine enough to reveal detailed structures in clouds.
The box length is long 
enough to incorporate most fluctuation powers. Greater $N$ (e.g. $2^{16}$) 
has also been tested with no difference found. The neutral hydrogen density 
$n_{HI}$ and the Ly$\alpha$ optical depth are calculated directly at each 
grid point according to our assumption of photo-ionization equilibrium
and Eq. 6.  Specifically,
the neutral fraction is 
\begin{equation}
f(T, J_{21}, n_e) = \frac{\alpha (T)}{\alpha (T) + \Gamma (T) + 
G_1 J_{21}/n_e},
\end{equation} 
where $G_1 = 2.83\times 10^{-12}$ for a UV spectrum with
power-law index -1.5, $\alpha
(T)$ is the recombination rate, $\Gamma (T)$ the collisional ionization
rate, and $n_e$ the number density of electrons (e.g. Black 1981). 
The detailed UV spectrum of Haardt \& Madau (1995) predicts
$G_1 = 3.2 \times 10^{-12}$, almost the same as the pure power law.

We create a reference density 
field by filtering $n_{HI} (y)$ with a Gaussian
function with dispersion $b_m$, with $b_m$ in the range 0 to 
$(2kT_m /m_p)^{\frac{1}{2}}$. Then we define each 
region between two successive minima in the reference field as an {\em 
absorber} or {\em cloud}, and each maximum between the minima as the central 
position of the absorber. The reference field is used only for the purpose 
of defining a {\em boundary}. It is not used even for defining the 
physical size of a cloud. Statistical properties of 
clouds that will be defined in Section 3.2 are always calculated directly 
from the original $n_{HI}$. With $b_m =0$ (``no smooth''), the reference 
field is identical
to $n_{HI}$ and we will select every feature in $n_{HI}$ as a separate
cloud. With $b_m = (2kT_m /m_p)^{\frac{1}{2}}$ (``maximal smooth''), density 
features on scales much smaller than the thermal broadening are erased
and the clouds identified match best to real Voigt lines in QSO spectra.
The reason for smoothing the reference field is 
twofold. First, we should not include any sparse enhancement over the local
background density profile of an absorber as a new absorber. And second, any 
fine structures that have been smoothed out by $b_m$ may be unobservable 
because of thermal broadening, so we should let the term ``cloud'', which 
usually implies ``one line'' in the spectrum, refer to just ``one region'' in 
space. We find that $b_m$ with ``maximal smooth'' can be applied 
to moderately resolved QSO spectra, while $b_m$ with ``no smooth''
may be applied to very highly resolved spectra like Keck observations.

The optical depth is calculated directly from $n_{HI}$ and has nothing to do 
with this selection procedure for identifying clouds. 

\section{Properties of the IGM and Ly$\alpha$ Clouds}

\subsection{Global evolution of mass and volume filling factors}

The LN field describes a density evolution in which mass moves slowly to 
high-density regions while more and more volume is left with low 
densities. Denote $\rho = n/n_0$ and $\sigma ^2$ the variance of the linear 
density; the fraction of volume $V(>\rho )$
occupied by densities above a given $\rho$ 
and the associated mass fraction $M(>\rho)$ are, respectively,
\begin{equation}
V (>\rho) = \int ^{\infty} _{\rho} p(\rho) d\rho
= \frac{1}{2} {\rm erfc}(\frac{\sigma}{2\sqrt{2}} + \frac{\ln(\rho)}
{\sqrt{2} \sigma}) ;
\end{equation}
\begin{equation}
M(>\rho) = \int ^{\infty} _{\rho} \rho p(\rho) d\rho
= \frac{1}{2} {\rm erfc} (-\frac{\sigma}{2\sqrt{2}} + \frac{\ln(\rho)}
{\sqrt{2} \sigma}) ; 
\end{equation}
where $p(\rho)$ is the one-point distribution function of $\rho$,
and the variance of $\rho$ is $\exp(\sigma ^2)-1$. We have 
plotted these functions at selected redshifts from $z=40$ to $z=0$ in Fig. 1 
for the CDM model. Fig. 2 shows the evolution of the volume $V(\rho <1)$ 
(solid line) occupied by densities less than the mean and their 
associated mass fraction $M(\rho <1) = 1-V(\rho <1)$ (dashed line), and the 
volume $V(\rho <0.5)$ and the associated mass $M(\rho <0.5)$.

\begin{figure}[h]
\vbox to3.6in{\rule{0pt}{3.6in}}
\includegraphics{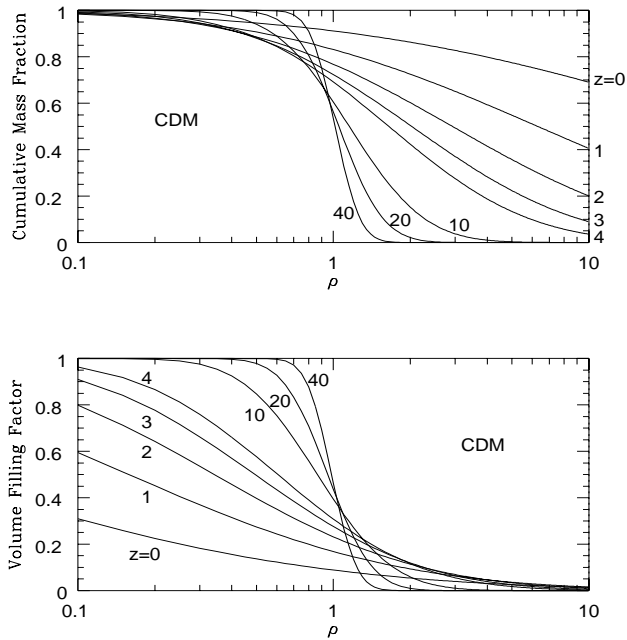}
\caption{Cumulative mass fractions and volume filling
factors from mass densities in the IGM above a given density threshold
$\rho \equiv \frac{n}{n_0}$ at
redshifts $z=40$, $z=20$, ..., to $z=0$ in the CDM model.
These curves are derived from our hypothesis that the probability
distribution function of the intergalactic gas density
is given by the lognormal
distribution.} \label{fig1}
\end{figure}

The different curves in Fig. 1 are determined uniquely by the standard
deviation $\sigma$, which is plotted in Fig. 3. The solid line
in the figure is for the CDM model and the dashed line for
the LCDM model. Some specific values for the CDM model are $\sigma = 0.87$
at $z=5$, 1.00 at $z=4$, 1.18 at $z=3$, 1.47 at $z=2$ and 1.90 at $z=1$.
For comparison, $\sigma = 1.0$ at $z=3$ in MCOR.
Using numerical simulations in a cold dark matter model with the power
spectrum 1.4 times smaller than ours, 
Carlberg \& Couchman (1989) obtained $\sigma = 0.744, 0.953, 1.39$ and 1.44
at $z=4.63, 2.80, 1.36$ and 0.81 for the linear IGM
(see Table 2 of Miralda-Escud\'e \& Ostriker 1990).  These values are
consistent with our calculations.

At $z=3$, the IGM in our model is just a median fluctuated
field in which both overdensities and underdensities have considerable
filling factors and are significant in appearance.
Yet there is still enough contrast to cause line-like absorption.
In the very early universe ($z>10$), half of the volume is occupied by
gas whose density is
below the mean, with about half of the total mass contained. 
As the fluctuations increase, mass moves into overdensities above the mean.
Fig. 1 shows that
at redshift 4, only 30\% of the mass is in the underdense regions, but this
mass occupies 70\% of the volume. At redshift 1, this becomes $\sim$ 20\% of 
the mass in $\sim$ 80\% of the volume. 

\begin{figure}[h]
\vbox to3.0in{\rule{0pt}{3.0in}}
\includegraphics{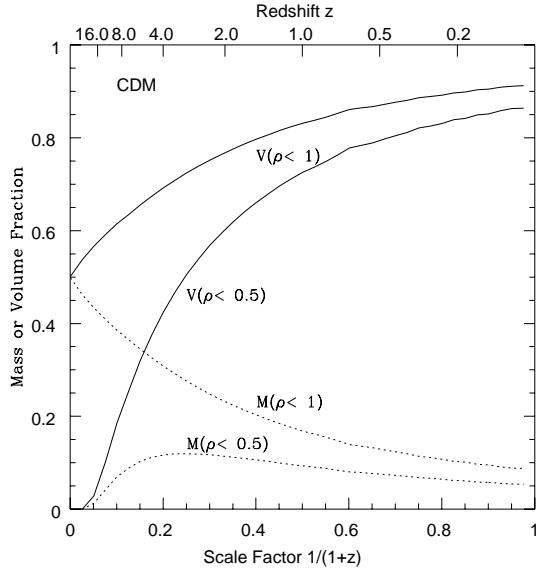}
\caption{The solid curves show the evolution
of volume filling factor vs. the scale factor $a = (1+z)^{-1}$
in the CDM model for
gas at densities $\rho \le 0.5$ or $\rho \le 1.0$. The dotted curves
show the mass fractions at these densities. \label{fig2}}
\end{figure}

\begin{figure}[h]
\vbox to3.0in{\rule{0pt}{3.0in}}
\includegraphics{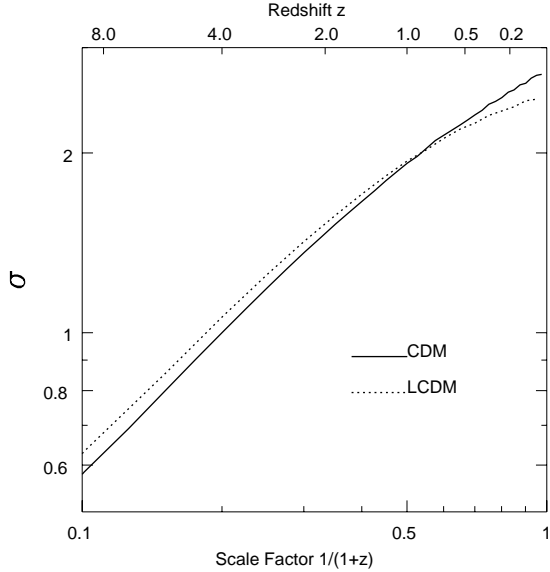}
\caption{The growth of fluctuations in the IGM
vs. the scale factor. The variance of the linear fluctuation is $\sigma ^2$,
while the variance of the density distribution of the LN model is
$\exp (\sigma ^2) -1$.
The evolution of $\sigma$ is shown as a solid curve for the CDM model,
and a dotted curve for the LCDM model. \label{fig3}}
\end{figure}

As we will see in \S 3.2, a typical Ly$\alpha$ cloud at $z=3$ with $N_{HI} = 
10^{13.5}$ cm$^{-2}$ will have a density approximately equal to the mean,
$\rho = 1$, and an $N_{HI} = 10^{12.0}$ cm$^{-2}$ absorber has $\rho = 0.2$; 
therefore the clouds between $10^{12.0}$ and $10^{13.5}$ cm$^{-2}$ are 
{\em underdense} ! 

We have made similar plots to those in Fig. 1 for the LCDM model, but the 
difference is almost invisible because the variance is so similar in the 
two models, as shown in Fig. 3. For example, $\sigma = 1.25$
at $z=3$, compared to 1.18 in the CDM model.
Therefore, we have not repeated the figures again. In fact, 
we find most of the statistical quantities are
identical in both the models, so they will be illustrated for the CDM 
model only. In this paper, if it is not stated
explicitly, the LCDM model may always be assumed to have 
essentially the same plots
as the CDM model.

\begin{figure}[h]
\vbox to5.0in{\rule{0pt}{5.0in}}
\includegraphics{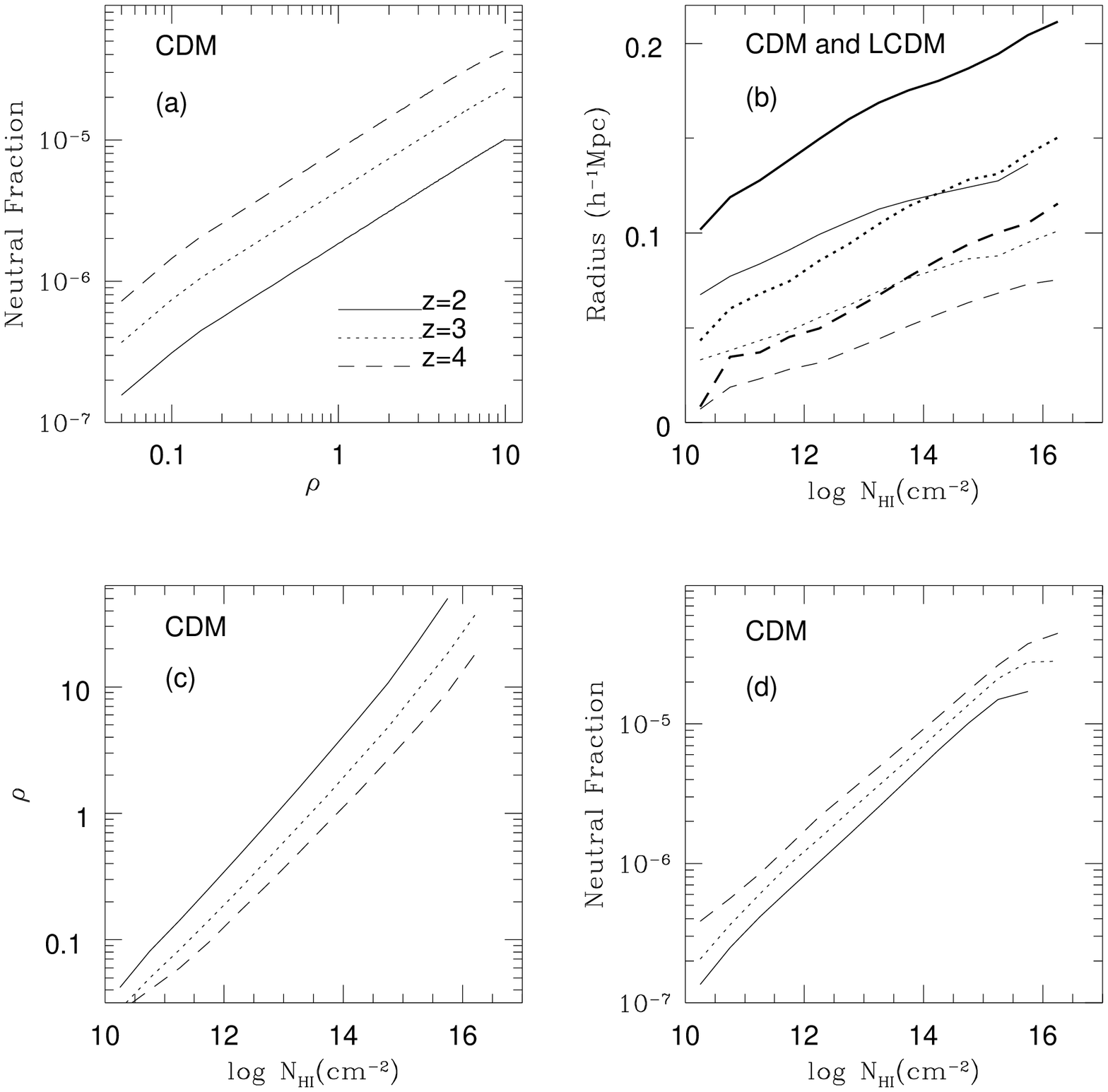}
\caption{({\bf a}) the neutral fraction vs. the density
$\rho$ in the IGM at three different redshifts : $z=2$ (solid), $z=3$
(dotted) and $z=4$ (dashed) in the CDM model. ({\bf b}) the physical
``radius'' (see text)
of Ly$\alpha$ absorbers at different column density. The light and heavy
curves are for the CDM and LCDM models, respectively. Figs. 4c and 4d show
the mean density and neutral fraction vs. the column density. \label{fig4}}
\end{figure}

\subsection{Photo-ionization of the IGM}

At temperatures of $10^3 - 10^5$ K, the only significant ionization 
process in the IGM is photo-ionization. The relation between the neutral 
fraction and the density is shown for the CDM model in Fig. 4a at 
redshift $z=2$ (solid line), 3 (dotted line) and 4 (dashed line) using 
$J_{21} = 0.18$ with $\Omega _b = 0.015h^{-2}$, or 
$J_{21} = 0.50$ with $\Omega _b = 0.025h^{-2}$. 
(The LCDM model gives the same results
with the same $J_{21}$.)

Let's define a few physical parameters of neutral hydrogen absorbers.

The column density $N_{HI}$ of an absorber is obtained by summing neutral 
densities in each pixel from the first minimum to the next minimum in the
reference field. The column density of {\em total} hydrogen
can be calculated in the same way. The ratio between the neutral and
total column densities defines the neutral fraction of the cloud. 

The mean density of a cloud is the density-weighted density in each pixel :
$<\rho > = \frac{\sum n_i ^2}{n_0 \sum n_i}$. 
Because in photoionization equilibrium
the neutral density is proportional to the square of the 
total density, this mean density can be calculated in another way, namely
the density deduced from the neutral fraction in Fig. 4a.
Note that both the column density and the mean mass density 
are defined independently of any detailed structure of the cloud, and also 
independent of the starting and ending points of the region outlined by 
the reference field; this is because most of the mass is contained sharply 
around the center of the density fluctuation.

The size $D$ of a cloud is defined to be the total column density divided by 
the mean density. This is just a convenient 
representation of the scale over which the absorber's 3-D density is 
typically measured. It may or may not be equal to the real diameter of the 
cloud even if it happens to be a sphere, but they should be
close to each other.

The physical ``radius'', which is taken to be half of the cloud size, is 
plotted in Fig. 4b with light curves for the CDM model and heavy curves for 
the LCDM models after averaging over 1000 random realizations. As a 
comparison, the physical Jeans radius, i.e. half of the Jeans length 
$\lambda _b$, is 0.13, 0.075, and 0.051 $h^{-1}$Mpc at $z=2, 3$ and 4 in 
the CDM model, and 0.21, 0.13, and 0.082 $h^{-1}$Mpc in the LCDM model. 
We note that the radius increases slowly with increasing column density. 
This is expected when clouds have no boundaries.
The radius for $N_{HI} = 10^{14}$ cm$^{-2}$ at $z=2$, which is 
110 $h^{-1}$kpc in the CDM model, is consistent with that measured for
the paired Q1343+266 A, B where the best fit value is around 90 $h^{-1}$kpc
(Bechtold et al. 1994; and Dinshaw et al. 1994). The radius in the LCDM model
is always larger than that in the CDM model because of the different 
cosmological geometry. If the universe is the flat LCDM model, the
detected radius of clouds will be larger than for the $\Omega = 1$ CDM 
model. Simulations of such paired spectra can be found in
Bi (1993) and Charlton et al. (1996).

The mean density of clouds is plotted vs. their neutral column
densities in Fig. 4c. As expected for 
photo-ionization, the
density is proportional to the square root of the column 
density. The column densities that correspond to the mean 
density
are $10^{12.9}$ cm$^{-2}$ at $z=2$, $10^{13.4}$ cm$^{-2}$ at $z=3$
and $10^{13.9}$ cm$^{-2}$ at $z=4$. It is perhaps surprising to 
see that these clouds are those that
appear most commonly in the Ly$\alpha$ forest,
yet they have just the mean density of the universe.
Fig. 4d plots the neutral fraction at various column densities. One can
see that high density
clouds have a higher neutral fraction. These curves can be compared
with Fig. 9b of MCOR. The agreement is good in general.

We have tabulated the total baryonic mass $\rho n_0 m_p D^3$, and the 
Jeans mass $n_0 m_p \lambda _b ^3$ in Table 1 in units of $\log _{10}
(M/M_{\odot})$. Fig. 5 shows the mass vs. column density at $z=2, 3$ 
and 4. Mass clumps of the constant fluctuations $\rho = 10$, 1, and 0.2
have been traced by the dashed arrows. The Jeans mass is marked as 
$M_J$ in the figure. The marked mass, $M_3$, is an example of a specific 
cloud found in our simulation, which we will discuss in detail in \S 5.2.

\begin{figure}[h]
\vbox to3.0in{\rule{0pt}{3.0in}}
\includegraphics{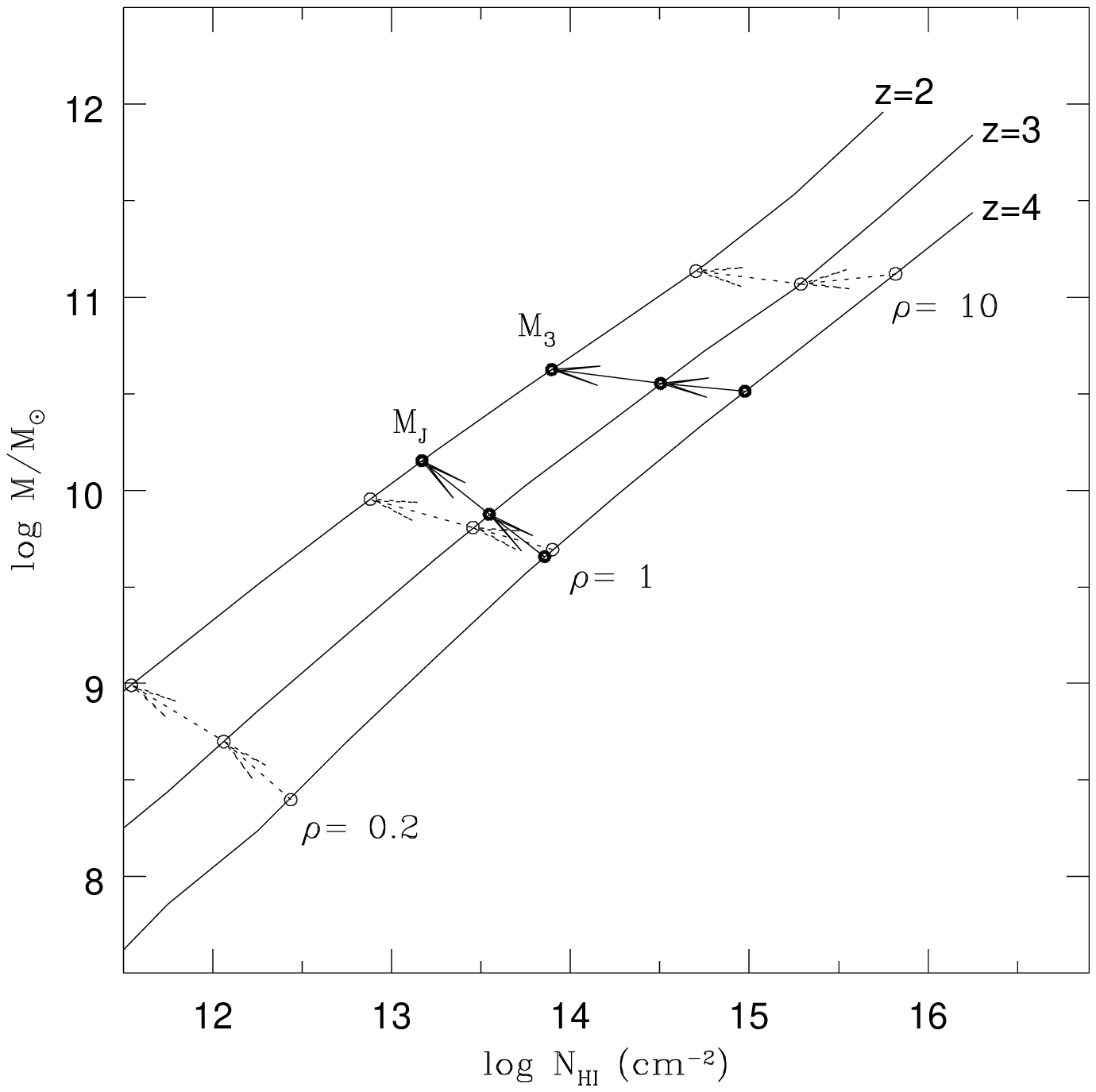}
\caption{Average baryonic mass of Ly$\alpha$ absorber vs.
column density at $z=2, 3$ and 4 in the CDM model. The absorbers at
the constant density $\rho = 0.2, 1.0 $ and 10 are traced as dashed arrows
and open circles. The Jeans mass $M_J$ is marked with solid points.
The solid points for $M_3$ trace the specific cloud No. 3 discussed
in \S 5.3. \label{fig5}}
\end{figure}

The meaning of the mass for low column density 
clouds should be interpreted with caution. 
They most likely originate from different impact 
parameters, and so represent only small parts of larger absorbers. 

\subsection{Structures within clouds}

The line-of-sight distributions of peculiar velocity and neutral density 
{\em within} clouds have been plotted (as the solid curves) vs.
the comoving separation from 
the center of cloud in the CDM and LCDM models in Fig. 6 at $z=3.0$. The 
successive curves with increasing peak values are mean profiles of all 
clouds between $10^{11} \to 10^{12}$ cm$^{-2}$, $10^{12} \to 10^{13}$ 
cm$^{-2}$, ..., $10^{15} \to 10^{16}$ cm$^{-2}$ correspondingly. 
The dashed curves show the same plots if we do not use the smoothed
reference field to select clouds, but count {\em any} maximum in
the density distribution as one cloud. The peculiar velocity shows that 
mass is falling toward the cloud centers in comoving space except for 
$N_{HI} < 10^{12}$ cm$^{-2}$. For clouds with $N_{HI} = 10^{13} - 10^{15}$ 
cm$^{-2}$, the bulk velocities are just comparable to the thermal 
broadening; for clouds with $N_{HI} = 10^{11} - 10^{13}$ cm$^{-2}$, they 
are negligible. Hence, most strong lines will have sharpened profiles due 
to the infall velocity. The maximal effect occurs at a separation of about 
0.2 to 0.4 $h^{-1}$Mpc, depending on the column density. The velocity is 
also larger in the LCDM model than in the CDM model at the same column 
density.

Below $10^{15.5}$ cm$^{-2}$, the peculiar velocities are smaller than the 
Hubble velocity at all distances; therefore, at $z=3$ the clouds are still 
expanding in proper coordinates, although in comoving space, they are 
condensing. The neutral column density, which is proportional to the square 
of the proper density of total mass,
decreases with cosmic time. Thus the same cloud will have a higher column
density at $z=4$ than at $z=2$. This will also be shown with examples
in \S 5.1. In this sense, 
none of the forest clouds is confined in proper space. Only for $N_{HI} \ge
10^{15.5}$ cm$^{-2}$, have we found stable or even collapsed
clouds in proper space
on the scale of $\sim 0.1 h^{-1}$Mpc. However, these clouds are very 
rare in the present simulation.
 
We find that the neutral hydrogen density profile of a cloud 
as illustrated in Fig. 6 is well 
described as 
\begin{equation}
n_{HI} (x) = \frac{N_{HI}}{x_0} \exp (-\frac{x}{x_0}),
\end{equation}
where $x_0 = 0.14 h^{-1}$Mpc for the CDM model, or 0.20 $h^{-1}$Mpc 
for the LCDM model. We note this $x_0$ is close to $x_b$ in Eq. 2;
therefore, the density declines exponentially outside the Jeans radius. 
If we take all maxima in the density distribution as separate clouds,
the cloud sizes and peculiar velocities become about half as large, but
their spatial structures (i.e. the dashed curves) are similar.

An important point is that not only the velocity but also the exponential 
density profile, which extends far away from the center 
of a cloud without a cut-off, 
will affect absorption line profiles. We call this {\em the absorption 
from density wings}. For instance, in the CDM model, a cloud with $N_{HI} 
= 10^{14.5}$ cm$^{-2}$ will have $n_{HI}  = 6\times 10^{-12}$ cm$^{-3}$ 
at 0.8 $h^{-1}$Mpc,
which is comparable to the peak density of a typical $N_{HI}= 10^{12.5}$ 
cm$^{-2}$ cloud. Therefore, the wing density at 0.8 $h^{-1}$Mpc, or 137 km 
s$^{-1}$ in velocity at $z=3$, is still contributing a noticeable 
absorption depth, since the $10^{12.5}$ cm$^{-2}$ column density is just 
detectable in observations. The resulting Ly$\alpha$ profile is thus much
{\em broader} than the pure Voigt profile of the cloud at the same column 
density but with all mass at its center. This effect exists in all 
clouds. As a short summary, the velocity sharpens lines at column density 
above $10^{13.5}$ cm$^{-2}$, but the wing absorption broadens lines in the 
same way for all absorbers. Hence, we have three kinds of departures
from the Voigt profile : the density wing, the peculiar velocity
and the line-blending effect. 

\begin{figure}[h]
\vbox to4.5in{\rule{0pt}{4.5in}}
\includegraphics{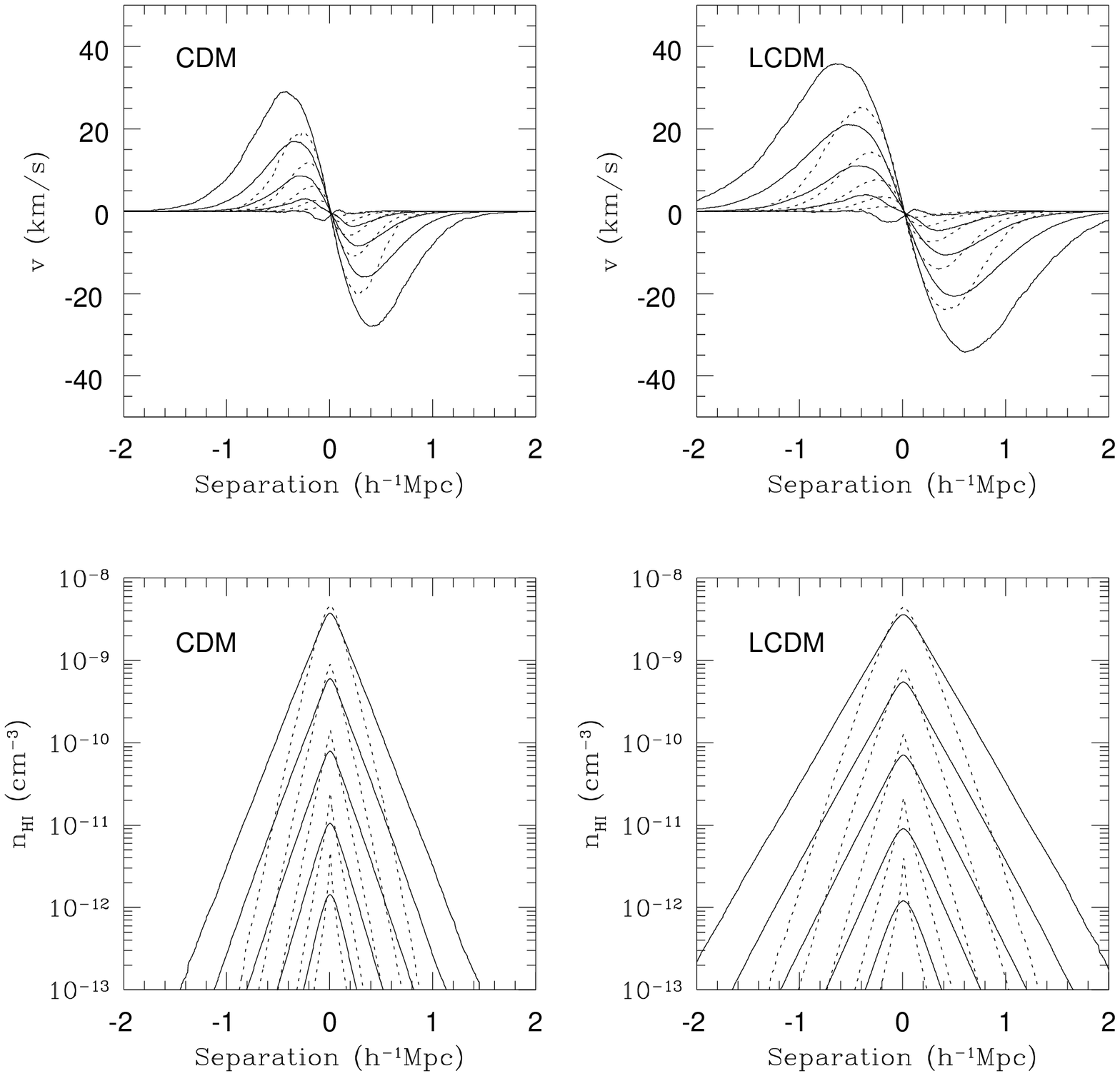}
\caption{Averaged density and peculiar velocity
profiles vs. comoving radius from the center of a cloud are plotted
as solid curves in both the CDM and LCDM models, using the selection
procedure in \S 3.1.
The curve with the highest amplitude corresponds to the cloud at $N_{HI}
= 10^{15.5}$ cm$^{-2}$ and the successive curves with lower amplitudes are at
$N_{HI} = 10^{14.5}, 10^{13.5}, 10^{12.5}$ and $10^{11.5}$ cm$^{-2}$,
respectively. Absorbers are selected according to the ``maximal smooth''
procedure. If we select {\em any} individual maxima in the line-of-sight
density distribution in the IGM as separate clouds, i.e. with the ``no
smooth'' procedure, the cloud dimensions
and the peculiar velocity become smaller, which are plotted as the dotted
curves. \label{fig6}}
\end{figure}

In real observations, the column density and
velocity fitted by multiple Voigt profiles will be significantly different
from the intrinsic column density
because of line-blending and fluctuations.
For an arbitrary absorber, it is 
almost impossible to say which effect -- the internal temperature, the 
peculiar velocity, or the density wing -- plays the most important role.
An exception is for {\em isolated} clouds with $N_{HI} \le 10^{13}$ 
cm$^{-2}$ : they are on the linear part of the curve-of-growth and thus
have unchanged $N_{HI}$; but their $b$ parameters will be significantly
boosted by the wing absorption. 
 
How the fitted parameters differ from actual parameters will deserve
a careful statistical study. Unfortunately, available Voigt fitting software
is very slow; we have not made extensive calculations for 
the present paper. A real 
Voigt fitting of a particular simulation sample will be illustrated in 
\S 5, however. 

\section{Global evolutions of HI Opacity and $dN/dz$}

We discuss the redshift evolution of the hydrogen Ly$\alpha$ opacity 
$\tau _{HI}$ and the number density of clouds $dN/dz$ in this section. 
The optical depth is proportional to the factor $\mu = 
\frac{(\Omega _b h^2)^2} 
{h J_{21}}$.  When $\Omega _b$ is fixed to the nominal primordial value, 
all uncertainties of the factor can be attributed to 
$J_{21}$. There is a number of $\tau _{HI}$ observations at $z\ge 1$
from which we need one at a particular redshift to normalize $J_{21}$.
The evolution of the opacity at other redshifts should then be automatically
generated by the IGM model. We choose $z=2.4$ because this is the mean 
redshift where the Hopkins Ultraviolet Telescope has measured the helium 
opacity toward HS1700+64 (DKZ) and for which 
there is also a good hydrogen Ly$\alpha$ spectrum observed by Keck
(Tytler 1995; Zheng et al. 1996).

\begin{figure}[h]
\vbox to3.0in{\rule{0pt}{3.0in}}
\includegraphics{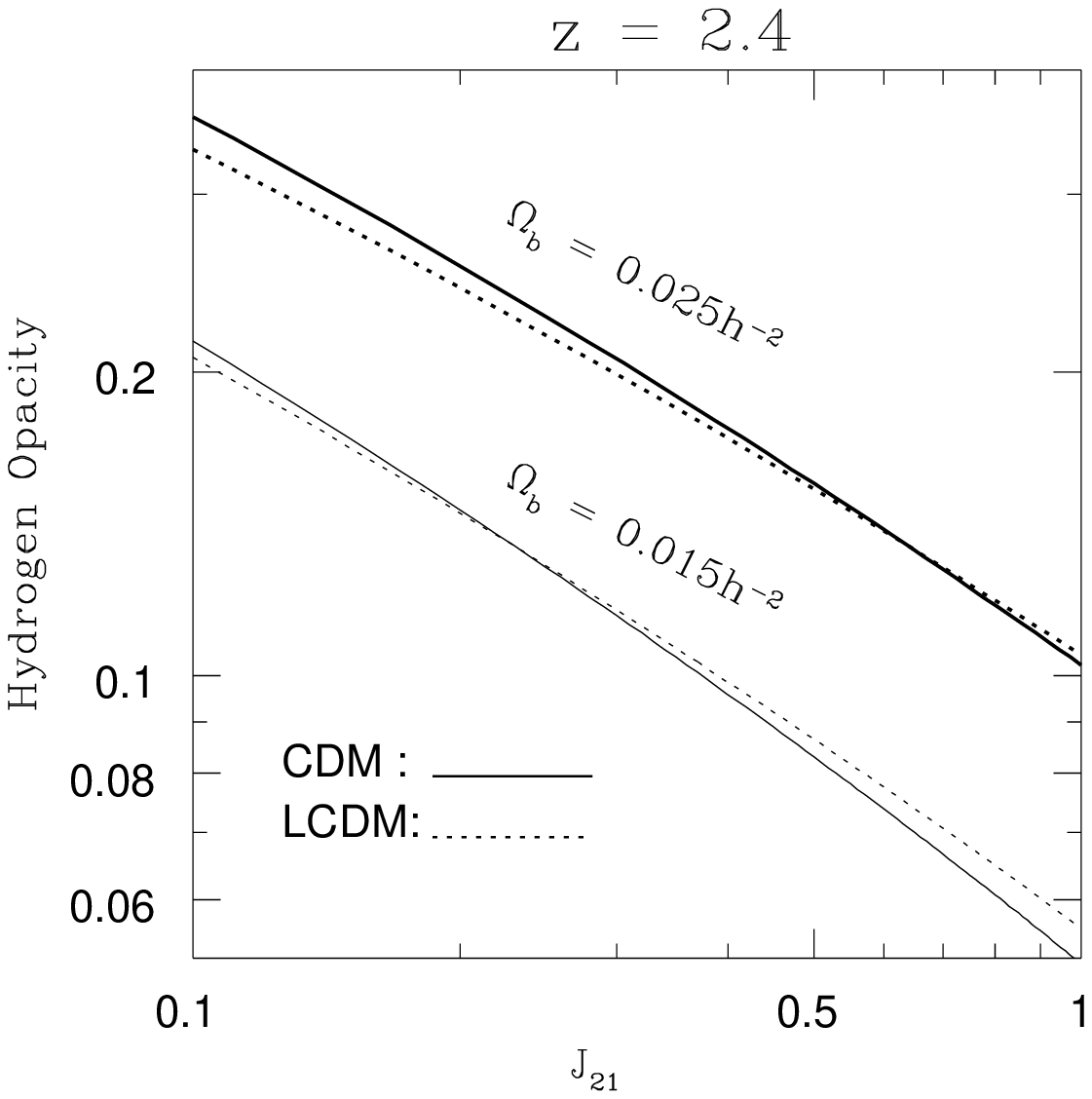}
\caption{The effective HI opacity at
$z=2.4$ vs. the UV background
flux $J_{21}$ for two different cosmological baryon densities :
$\Omega _b = 0.025h^{-2}$ and $\Omega _b = 0.015h^{-2}$. The solid and
dashed curves are for the CDM and LCDM models, respectively. \label{fig7}}
\end{figure}

The average Ly$\alpha$ opacity at $z=2.4$ is 0.153 according to the recent 
measurement of Dobrzycki \& Bechtold (1996). The opacity for HS1700+64
is $0.22\pm 0.01$ (Davidsen et al. 1996). The difference presumably comes
from random statistics, i.e. fluctuations in the number of intercepted
clouds due either to variation in the density or the UV background, or both.
The global value $\tau _{HI} =  0.153$ will be adopted in the following
discussion. When comparing synthetic samples directly to the Keck spectrum
of HS1700+64 in \S 5, we will use 0.22.
Fig. 7 plots $\tau _{eff}$ versus $J_{21}$ at $z=2.4$, where 
\begin{equation}
\tau _{eff} = -\ln <\frac{I_{obs}}{I_0}>
\end{equation}
is the effective opacity measured over a sufficiently large spectral 
range (243\AA \ in our simulation). Two values of
$\Omega _b$, $0.025h^{-2}$ and
$0.015h^{-2}$, are illustrated. The UV flux, $J_{21} = 0.50$ or 0.18 
that we have adopted in the last section has been selected for
$\Omega _b = 0.025h^{-2}$ or $0.015h^{-2}$, respectively, because the 
resulting HI opacity is 0.155 in both the models.

The UV flux $J_{21} = 0.18$ 
is lower than what is typically inferred from the QSO 
proximity effect, $J_{21} = 10^{\pm 0.5}$ (Bajtlik, Duncan \& Ostriker 
1988; Lu, Wolfe \& Turnshek 1991; Bechtold 1994).  However, it is 
consistent with the results of hydrodynamical simulations (e.g. Cen et al 
1994; MCOR; Zhang et al. 1995; Hernquist et al. 1996). 
The flux $J_{21} = 0.5$ is the value obtained for the observed quasars at
$z=2 - 3$ (Haardt \& Madau 1996). A recent study of the proximity
effect suggests $J_{21} = 1.0 ^{+0.5} _{-0.3}$ (Cooke, Espey \& Carswell
1996).

\begin{figure}[h]
\vbox to3.0in{\rule{0pt}{3.0in}}
\includegraphics{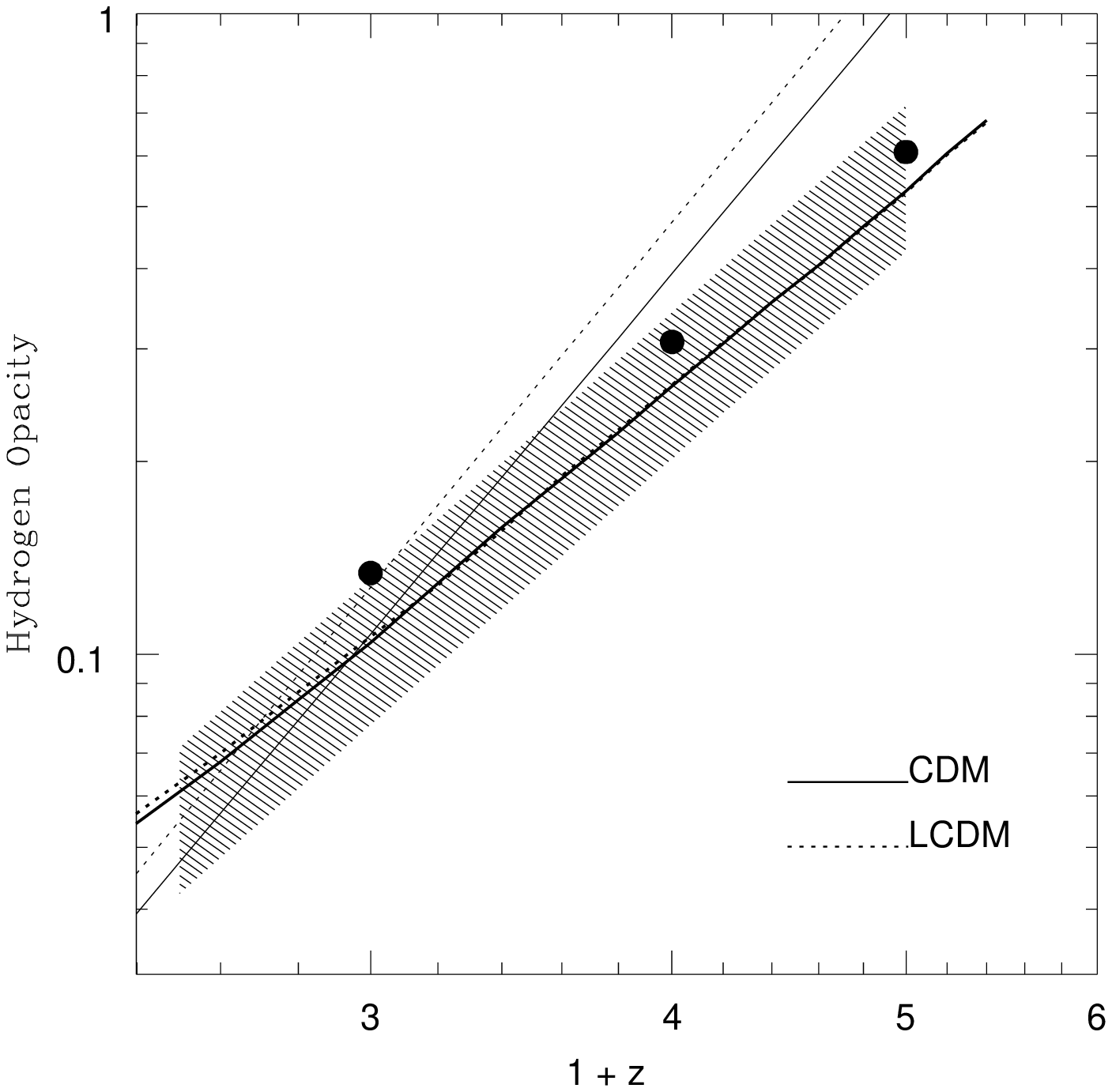}
\caption{Evolution of the effective
hydrogen opacity from redshift
4 to redshift 1.6 in the CDM (heavy solid line) and LCDM (heavy
dotted line) models, both
of which can be fitted by the power law $\tau \propto (1+z)^{3.1}$. The
ionizing background intensity is assumed to be constant. The
corresponding light-weight curves, $\tau \propto (1+z)^{4.5}$,
are what is expected for a uniform IGM with the same mean
density. The shaded
region is the observational curve $\tau _{eff} = 0.153
(\frac{1+z}{3.4})^{3.33}$ with 30\% variations from Dobrzycki \& Bechtold
(1996). The three solid points are from MCOR's simulation. \label{fig8}}
\end{figure}

If the IGM is uniform and $J_{21}$ is constant with $z$, the opacity 
will be proportional to $(1+z)^{4.5}$
(Jenkins \& Ostriker 1991). In a clumpy IGM, there are two opposing effects 
due to
clumping. First, the neutral density, which is proportional to the square 
of the total density (in the isothermal case), is increased by the clumping 
factor, and so is the opacity. Second, clouds of $N_{HI} \ge 10^{14}$ 
cm$^{-2}$ will be saturated in the curve-of-growth, so mass clustered into 
those clouds contributes much less to the absorption. The 
results of our calculation are plotted in Fig. 8, where the observations 
of Dobrzycki \& Bechtold (1996),
\begin{equation}
\tau _{HI} (z) = 0.153 \times (\frac{1+z}{3.4}) ^{3.33},
\end{equation}
are illustrated (with $\sim$ 30\% variations) as the shaded region. The 
two heavy curves are for the CDM (solid) and LCDM (dotted) models, 
respectively, which are best 
fitted by $\tau = 0.26 [(1+z)/4] ^{1+\gamma}$ 
with index $\gamma = 2.1$. 
The corresponding light curves are for the case of a
uniform IGM with the same mean density. The three solid
points are the simulation results of MCOR.
The two effects of clustering compensate with each other at redshift 2.0,
giving an effective opacity which is the same for the clumpy IGM as it
would be for a uniform IGM of the same mean density.
At lower redshift, the first effect dominates but at higher redshift, the 
second effects dominates. Consequently we obtain more opacity at low redshift
and less opacity at high redshift than what is expected for a uniform
IGM. The power law of
the evolution is thus changed from index 3.5 to 2.1 in our simulation.
The agreement between the observations and the model is acceptable
with a constant UV flux for $z=2 - 4$; 
evolution of $J_{21}$ over this range of redshifts
is therefore not required.

Most of the hydrogen opacity originates from clouds
of column density greater than $10^{13}$ cm$^{-2}$,
with total mass density close to or above the universal mean density.
Hence they contain most of the mass too. To show this
fact more clearly, we have plotted the cumulative mass fraction for clouds 
{\em above} a given column density in Fig. 9a for $z = 2, 3$ and 4 in the 
CDM model. The cumulative HI opacities are plotted in Fig. 9b, and the 
cumulative number of clouds per unit redshift are 
shown in Fig. 9c. There is very little
contribution to the HI opacity from clouds below $10^{13}$ cm$^{-2}$
because of their negligible total mass, 
which is quite consistent with the studies where absorbers are treated 
as discrete clouds. The mass and the HI 
opacity are proportional to each other in the model. However,
there is not a one-to-one correlation
between the total number of clouds and the opacity.
While only a few clouds are found below $10^{13}$ cm$^{-2}$ at $z=4$, 
there exists a large number of low column density clouds at $z=2, 3$ 
in Fig. 9c.  Fig. 9d is a plot of the differential column
density distribution. MCOR made the same plot
as their Fig. 15b. The straight line is a power law of index
-0.5 that corresponds to $\beta = -1.5$
in the traditional $d^2N/dzdN_{HI} \propto N_{HI} ^{-\beta}$ function. 

The evolution of the column density distribution versus
redshift is apparent; the difference cannot be explained simply
by a factor in the amplitude. The three curves at $z=2, 3$ and 4 can
be shifted to overlap each other in the x-axis. This shows an important
fact : that the evolution of the
Ly$\alpha$ forest is not due to creating or destroying of new clouds,
but due to a {\em change of column density in the same clouds}. Because of
the expansion of the universe, clouds have smaller mass density (and neutral
density) at low redshifts than at high redshifts.
An interesting conclusion is that there are more $10^{12-13}$
cm$^{-2}$ absorbers at $z=2$ than at $z=4$. This may be tested by
future Keck observations. See Fig. 15b of MCOR for a similar conclusion.
 
\clearpage

\begin{figure}[h]
\vbox to4.7in{\rule{0pt}{4.7in}}
\includegraphics{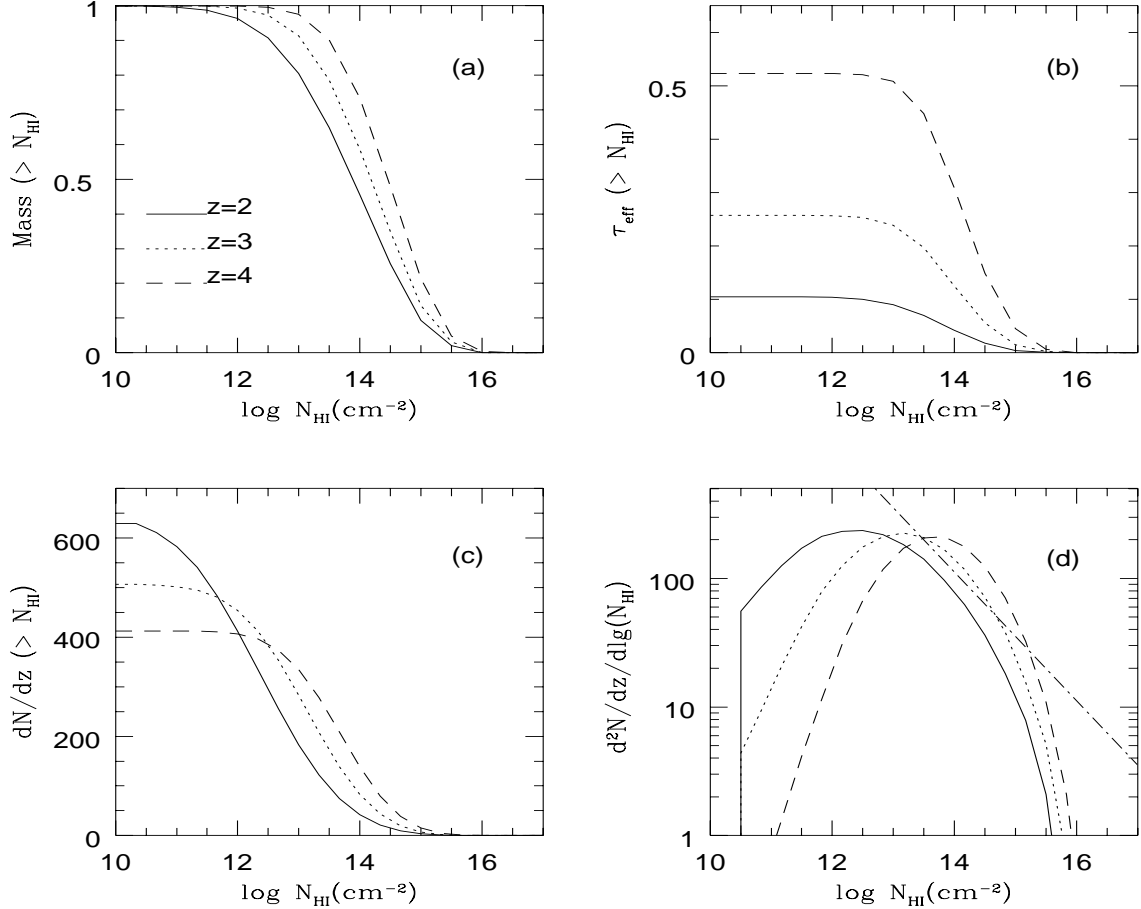}
\caption{The cumulative mass fraction, HI opacity, and number
of clouds per redshift above a given column density are illustrated in
Figs. 9a, 9b and 9c, respectively, for three redshifts
$z=2$ (solid), $z=3$ (dotted) and $z=4$ (dashed) in the CDM model.
Fig. 9d is the number of clouds per redshift per $\log N_{HI}$. The dot-dashed
curve is a power law of index -0.5. \label{fig9}}
\end{figure}

The column densities here are intrinsic values, i.e., calculated directly
from the neutral hydrogen distribution. In real observations, Ly$\alpha$
lines are blended so that both the number of lines and the column 
densities in a blended absorption spectrum depend on the resolution and 
signal-to-noise ratio of the observation. This ``blending effect''
will introduce a correction factor between the number of intrinsic clouds
and the number of observational lines fitted according to various
Voigt profiles. The best way to solve this
difficulty is to fit simulated spectra in the same way as observations.
However, the well-known Voigt fitting software, VPFIT, is too slow to be
useful because the simulation data set is huge. In section \S 5.1,
we have run VPFIT to one of the simulation samples at $z=2.4$
and have found that the number of intrinsic clouds is about 1.70 times 
larger than the number of lines found in the simulated spectrum and fitted 
by Voigt profiles in $12.5 \le \log N_{HI} \le 15.0$. Fig. 10 shows the 
column density distribution $f(N)$ in units of counts per column density 
per $X$, where $X = 0.5(1+z)^2-1$. The relation between $f(N)$ and 
$\frac{d^2N}{dzdN}$, the counts per column density per redshift, is
$f(N) = \frac{d^2N}{dzdN} / (1+z)$. The solid data points are from 
the raw data of 4 Keck spectra (Hu et al. 1995), and the error bars are from
the earlier statistical results (Petitjean et al. 1993). Two power
laws are plotted. The dot-dashed straight line is the power law of index
-1.46 and the short dashed line of index -1.80, which fit to the 
observational data below $10^{14}$ cm$^{-2}$ and above $10^{14}$ cm$^{-2}$
respectively. The light solid curve is from the CDM model using 
``maximal smooth'' (the reference field is smoothed on the scale of
the thermal dispersion), and the heavy long-dashed curve using ``no smooth''
(every feature in the neutral hydrogen distribution is regarded as a separate 
cloud). The corresponding curves in the LCDM model are almost the same.
The factor 1.70 mentioned above has been applied to correct the number of
clouds in the simulation to the number of lines expected.

\begin{figure}[h]
\vbox to3.0in{\rule{0pt}{3.0in}}
\includegraphics{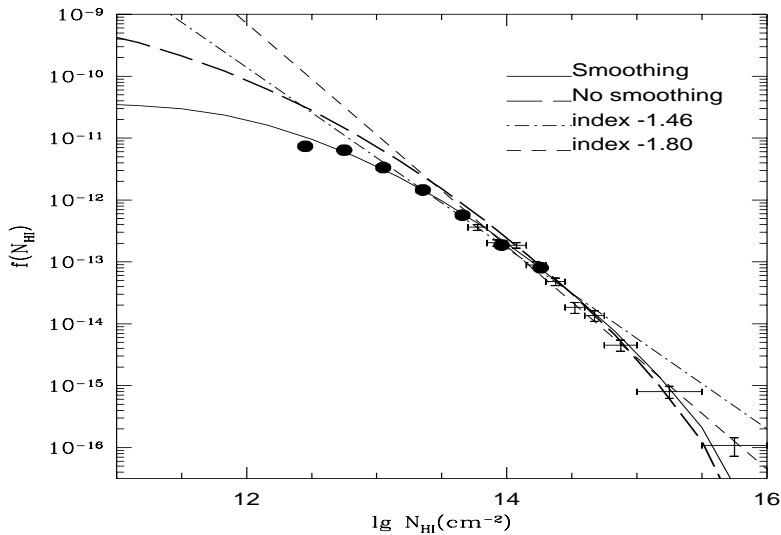}
\caption{Probability distribution of column density
defined as counts per column density per $X$ with $X = 0.5 (1+z)^2 -1$.
The original observations are plotted as dot points
from Hu et al. (1995) and error bars from Petitjean et al. (1993).
The dot-dashed and short dashed straight lines are power laws of
index -1.46 or -1.80, respectively. The light solid curve is from
the CDM model using the ``maximal smooth'' procedure in \S 3.1.
The heavy long dashed curve is without smoothing (the ``no smooth''
procedure). The factor of 1.70 has been taken into account relating
the number of intrinsic clouds and the number of clouds
derived in the Voigt fit to the simulated spectra (see \S 5.1). \label{fig10}}
\end{figure}

The CDM model matches the Keck result quite well in the range $13.0 \le \log
N_{HI} \le 14.0$. Below $10^{13.0}$, some Ly$\alpha$ lines would be missed even
in the Keck data. Our smoothing procedure in the
cloud selection has underestimated the counts too, 
apparently because such small features are easier to be erased. 
The distribution $f(N)$ from clouds without smoothing can fit the 
``original''
counts of clouds at $\log N_{HI} \le 13.0$, which is suggested to fit
the power law with $\beta = -1.46$
in the figure (Hu et al. 1995). At somewhat higher column
densities the distribution steepens, giving $\beta = -1.80$ over the range
$14.0 \le \log N_{HI} \le 15.5$, and matching the results of 
Petitjean et al. (1993). At $\log N_{HI} \ge 16.0$, we find almost no clouds
in the simulation, indicating that these 
absorption systems have other physical origins
than do the ordinary Ly$\alpha$ forest clouds.
We associate these Lyman limit systems and the even higher column density 
damped Ly$\alpha$ systems with collapsed objects, such as galactic disks.
These much rarer high column density systems are clearly highly non-linear
and would not be expected to be produced in our calculation.

\begin{figure}[h]
\vbox to3.0in{\rule{0pt}{3.0in}}
\includegraphics{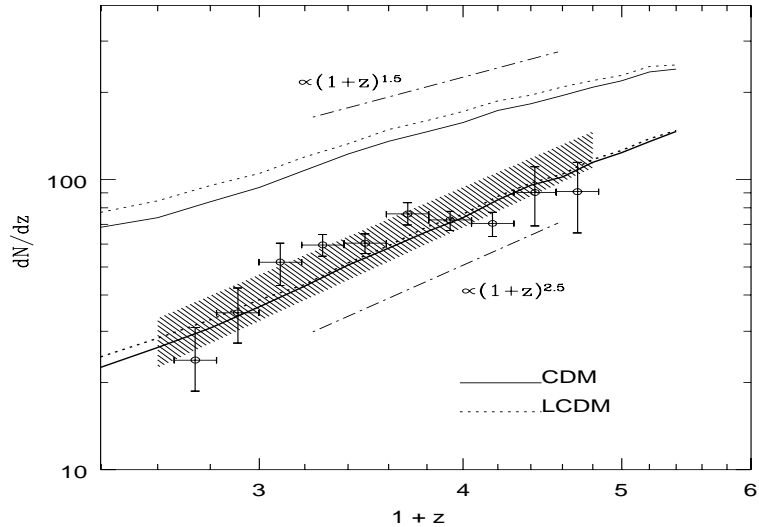}
\caption{$dN/dz$ for $W_0 \ge 0.32\AA$. The data points are
from Bechtold (1994) and the shaded region is the fit formula with
30\% variations from Lu, Wolfe \& Turnshek (1991). The heavy solid
and dotted curves are for the CDM and LCDM models, respectively, both of
which are best fitted by a power law $\propto (1+z)^{2.5}$. The light curves
are $dN/dz$ for Ly$\alpha$ lines selected with their
minimum transmitted fluxes below 0.5, which are best fitted by $\propto
(1+z)^{1.5}$. \label{fig11}}
\end{figure}

For each simulated spectrum, lines are selected from two successive 
maxima and their equivalent widths are calculated directly.
The resulting 
$dN/dz$ for strong lines at $W\ge 0.32\AA$ are plotted in Fig. 11 as the 
heavy solid curve (CDM) or the heavy dotted curve 
(LCDM).
The evolution is fitted best by $dN/dz = 75 [(1+z)/4] ^{2.5}$. 
This index is different from the $\gamma$ found from $\tau _{HI}$, 
which was 2.1, due to the evolution of the column density distribution in 
Fig. 9d, where high column densities are more common at large 
redshifts and have stronger saturation. 
We should, however, compare these counts with observations not 
polluted by the proximity effect. The
data points with error bars are from Bechtold (1994). The shaded region
is the fitting formula for the ``old-subset'' sample in 
Lu et al. (1991) (note that this sample is selected
at $W \ge 0.36\AA$). 
Lines can be selected not only by the equivalent width, but also
by the central minima of the line profiles if the profiles are resolved
as in Keck HIRES data. The two thin curves in the 
figure are for lines with the minimum transmitted flux less than 0.5; the 
solid curve is for the CDM model and the dotted curve for the LCDM model. 
The best fit for this definition is $dN/dz \propto (1+z)^{1.5}$.
There are no data published yet for comparison with these curves.

\section{Similarity between Simulations and Observations}

\subsection {A real synthetic sample}

Let us compare directly the Keck spectrum of HS1700+64 and synthetic 
spectra in this section. The data were obtained by D. Tytler and are 
discussed in detail in Zheng et al. (1996).
As we discussed before, the Gunn-Peterson
opacity of HS1700+64, $\tau _{HI} = 0.22\pm 0.01$, is greater than the
global average value 0.153 at $z=2.4$. Therefore we use $J_{21} = 0.1$ with
$\Omega _b = 0.015h^{-2}$ to simulate
absorption spectra for this specific quasar, which produces $\tau _{HI} =
0.21$ in both the dark matter models. 
 
The Keck spectrum in the redshift range $z=2.3$ to
$z=2.5$ is plotted in the upper panel in Fig. 12. A careful check 
reveals many line profiles are asymmetric or contain shallow non-Voigt
wings (such as the ones at 4033\AA , 4127\AA , 4207\AA \ and 4251\AA ), which 
reflect typically absorptions of asymmetric mass clumps in the diffuse 
IGM model. Particular groups of lines, such as those regions at 4070\AA \ to 
4077\AA , 4139\AA \ to 4143\AA \ and 4216\AA \ to 4222\AA , would require a 
considerable number of weak lines to explain, but they can be easily 
produced by mass clusterings instead of individual discrete clouds.
 
A synthetic sample in the CDM model is plotted below the Keck
data (the corresponding sample in the LCDM model is almost indistinguishable
and thus has not been plotted). The Keck data were obtained at a resolution 
of 3 km s$^{-1}$, so we can simply neglect the line-spread-function in the 
simulation. Gaussian noise of zero mean and variance 0.04 has been added to 
each pixel in the simulated spectrum to match the observations. 
The similarity between the simulation and the real data is very good. 

The simulation spectrum is analyzed using an accurate (but slow)
spectral fitting program called ``VPFIT'' by R. F. Carswell et al.. 
In Table 2, we list all lines with positions, $b$ parameters and column
densities. Lines with $N_{HI} \le 10^{12.5}$ cm$^{-2}$ are uncertain
and are not included in the table. The vertical bars in Fig. 12 mark the
position of the lines in the table.
There are several differences between the original,
intrinsic absorbers and the lines selected by ``VPFIT''.
1) Some intrinsic absorbers have been blended together to form {\em
one} strong line, e.g. three absorbers with total intrinsic
column density $3\times 10^{14}$ cm$^{-2}$ are blended to one strong
feature at 4092\AA \ that has the fit column density $\log N_{HI} = 17.0.$ 
In real observations, one usually needs associated metal lines to fix
the number of sub-components.
2) A few absorbers with broad density wings have to be fitted by 
{\em two} Voigt lines, one with a small $b$ for the main part of
the profile and the other with a large $b$ for the absorption wing. 
The absorber with
intrinsic $N_{HI} = 7.5\times 10^{13}$ at 4058\AA, \ for example, is 
split into two lines with a total $N_{HI} = 2.7\times 10^{14}$. 
3) Many weak features which can be marginally
identified by eye in the synthetic spectrum are actually associated 
with small mass clumps, not noise. This would suggest that similar features
in the Keck data may be real too.

\begin{figure}[h]
\vbox to4.8in{\rule{0pt}{4.8in}}
\includegraphics{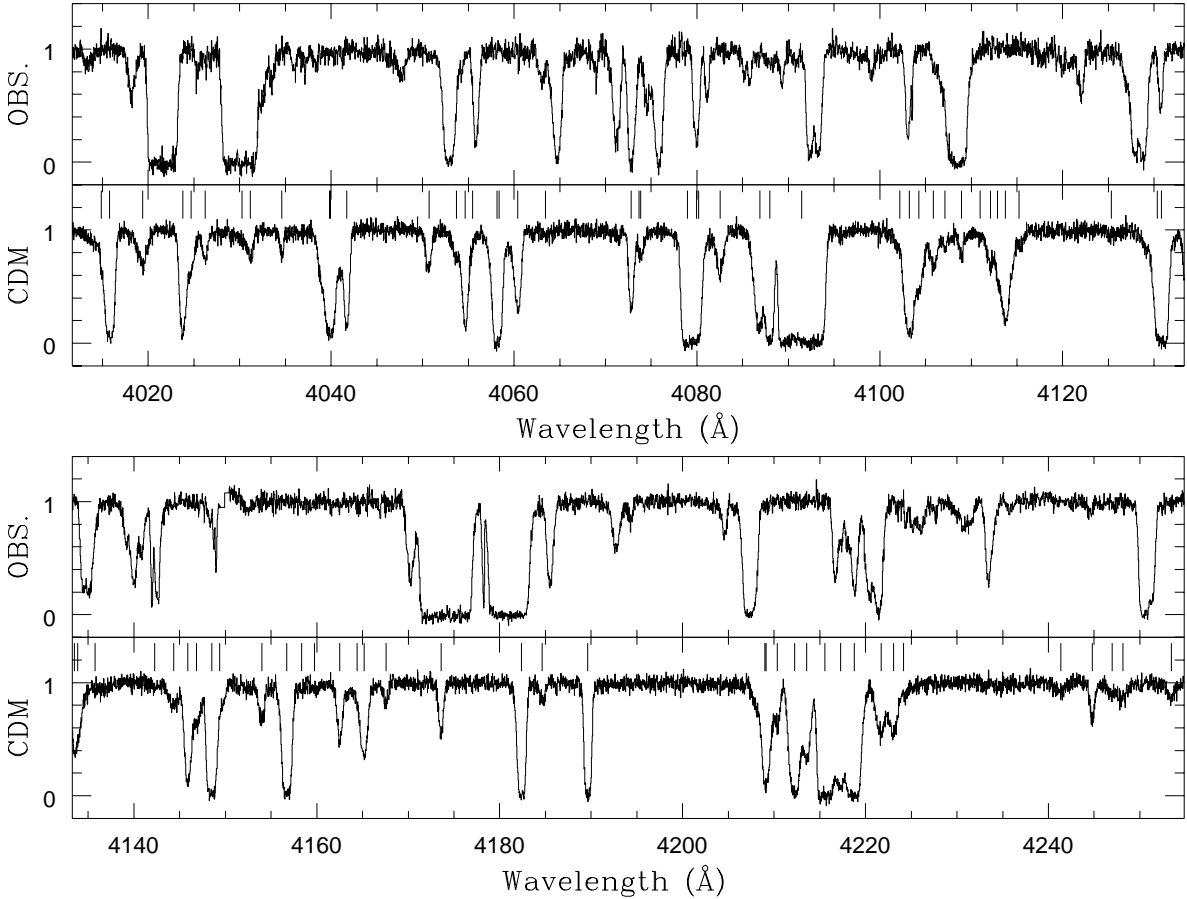}
\caption{The Keck spectrum of HS1700+64 in the redshift range
2.3 to 2.5, and a simulated spectrum in the CDM model. All lines with
Voigt-fit
column densities exceeding $10^{12.5}$ cm$^{-2}$ are marked as vertical
bars in the synthetic spectrum. \label{fig12}}
\end{figure}

As a summary,
the total number of lines with $\log N_{HI} \ge 12.5$ in Table 2 is 79, 
while the number of
intrinsic clouds is 135 in the same column density range. 
So the intrinsic number is approximately 1.70 time
greater than the number from the VPFIT software.

The histograms of the fitted $b$ parameter are shown in Fig. 13  as
the light histogram. The thick-line histogram is from the observation
of Carswell (1989). The shape of the simulated distribution is 
quite consistent with the observation (see also Hunstead 1988; 
Press \& Rybicki 1993). Both the distributions show that most of the 
lines have $b=20$ to 40 km s$^{-1}$ and that an extensive tail exists 
between 40 to 100 km s$^{-1}$. 

\clearpage
\begin{figure}[h]
\vbox to3.0in{\rule{0pt}{3.0in}}
\includegraphics{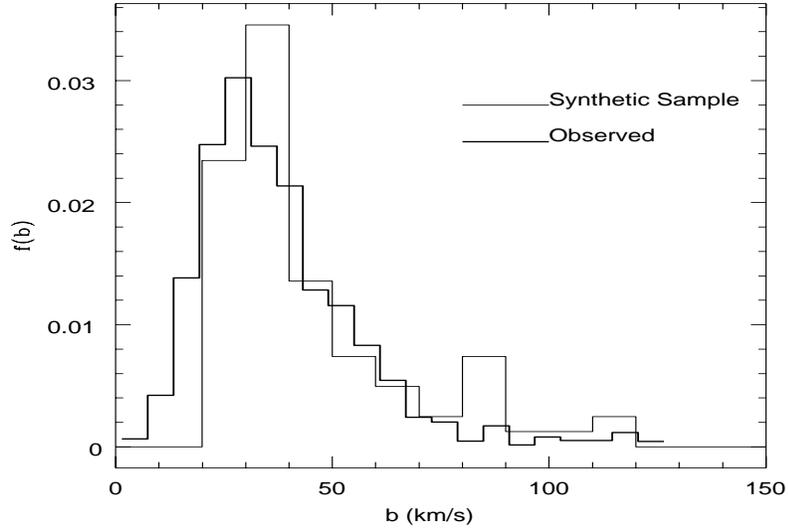}
\caption{The probability distribution of the $b$ parameters
of the Voigt-fit Ly$\alpha$ lines in our simulated spectrum (Fig. 12
and Table 2) is shown as the light histogram.
The heavy histogram is from observations (Carswell 1989). \label{fig13}}
\end{figure}

\subsection{Evolution of specific clouds}

Consider the IGM distribution in the box of 5 $h^{-1}$Mpc around $z=2.4$
in Fig. 12. The density $\rho$, peculiar velocity $v$ and absorption spectrum
are plotted in Fig. 14 for the same IGM mass distribution at
three different redshifts : $z=2$, 3, and 4.
The solid lines in the $\rho$ panels are the normalized density distribution
and the dashed lines the ``maximal smoothed'' reference field used in
the line identification.

We trace specifically 6 intrinsic absorbers
in the $\rho$ panel at $z=4$, which are marked as vertical bars.
The absorbers are
selected according to our selection procedure in \S 3.2.  Their intrinsic
column densities are listed in Table 3 as $\log N_{int}$.
The vertical bars in the
``Spectrum'' panels mark the fit absorption lines which correspond to
the intrinsic absorbers. The shifts of the
line positions are due to the peculiar velocity.
The dotted curves are the restored spectra
from the fit lines. The fit column density $\log N_{fit}$
and $b_{fit}$ parameters are also listed in Table 3. 
Note that the shape of the velocity
profile around the strongest line, line 3, agrees well with the globally 
averaged velocity profile in Fig. 6. We see that mass is falling from the 
two sides toward the line center in comoving space.

At redshift 4,
absorbers 1, 2 and 3 are seriously blended in both the $\rho$ distribution
and the spectrum, so it is not surprising that their fit column densities
are different from the intrinsic values. Yet the sum of the intrinsic
column densities, $10^{15.34}$ cm$^{-2}$, is comparable to the sum of the
fit column densities $10^{15.50}$. Absorbers 5 and 6 are relatively 
separated from the blended region, so their
$N_{int}$ are approximately equal to the fit $N_{fit}$.

\begin{figure}[h]
\vbox to4.8in{\rule{0pt}{4.8in}}
\includegraphics{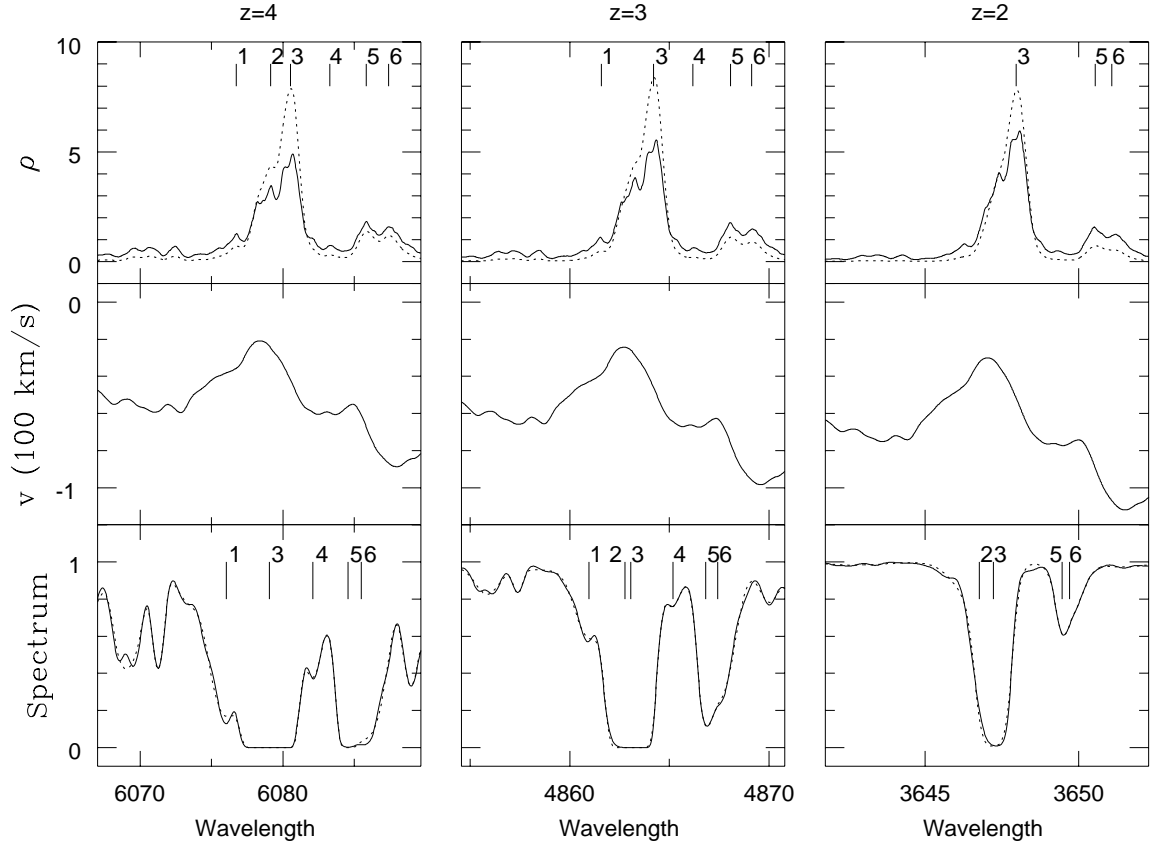}
\caption{Evolution of the IGM in the comoving
box 5$h^{-1}$Mpc around 4133\AA \ at $z=2.4$ in Fig. 12
at three different redshifts, $z=4$, $z=3$ and $z=2$. The upper
panels show the density distribution (solid lines) and
the ``maximal smoothed'' reference field (dashed lines)
where six intrinsic absorbers
are marked and discussed in the paper.
The middle panels show the peculiar velocity.
The lower panels are the absorption spectra. The marked numbers
are lines from the Voigt-fit and the dashed curves are the restored spectra
from these lines. \label{fig14}}
\end{figure}

At $z=3$ and $z=2$, the density fluctuation grows. Overdense
regions, such as absorber 3, are becoming even denser, and small density 
clumps, such as absorbers 1 and 4, are being destroyed or merged into 
large absorbers. Because the mean cosmological density is decreased, all 
the clouds have evolved to smaller column densities. Their $b$ 
parameters are usually smaller at lower redshifts too. Of particular 
interest is the cloud marked Number 2. At $z=4$, it has been selected by 
our ``maximal smooth'' identification. Because its absorption is very 
strong and blended with cloud 3, we only detect one spectral line (marked 
Number 3) from the two absorbers. At $z=3$, cloud 2 can no longer be 
regarded as a separate cloud in the ``maximal smooth'' identification
(although it is still present
in the ``no smooth'' identification, of course); however,
the asymmetric profile of the merged absorption does require 
another component in the Voigt fitting, which is marked as line 2. 
It is noticed that cloud 2 does not match line 2 in either position or column 
density. This shows that the Voigt profile fitting is not efficient
in revealing intrinsic structures if the IGM
is not in discrete clouds, but is continuously distributed in space
as we suggested here.

We also notice that the three underdense clumps around 6070 \AA \ at $z=4$
are disappearing in the spectra from $z=4 $ to $z=2$, which can be 
understood again by the two effects :
their density contrasts become even lower due to clustering, and the
average neutral hydrogen density is lower due to the expansion of the 
universe.

\begin{figure}[h]
\vbox to3.0in{\rule{0pt}{3.0in}}
\includegraphics{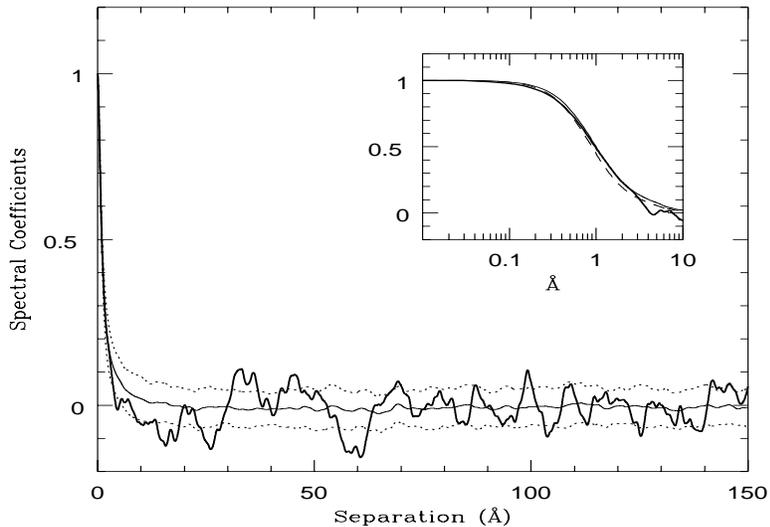}
\caption{Averaged spectral coefficients vs. separations for
1000 synthetic samples similar to that in Fig. 12. The two dotted curves
show the $1\sigma$ error bars in the simulations. The graph in the insert is
the same plot but on very small separations where we have also plotted
the LCDM model as the dashed curve. The heavy solid curve is from the
Keck observation of HS 1700+64. \label{fig15}}
\end{figure}

\subsection{Autocorrelation coefficients}

We can view an absorption spectrum as a time series.
To see how the spectrum $f(k), k=1, 2, ..., N$, is correlated at the 
separation of $l$ pixels, the following autocorrelation coefficients
can be used (Bi 1993; MCOR) : 
\begin{equation}
r_c (l) = \frac{1}{N\sigma ^2} \sum ^{N-l} _{k=1} [f(k) - <f>]
[f(k + l) - <f>], \ \ \ l = 0, 1, 2, ..., N-1;
\end{equation}
where 
\[
<f> = \frac{1}{N} \sum _{k=1} ^{N} f(k), \ \ \
\sigma ^2 = \frac{1}{N} \sum _{k=1} ^{N} (f(k) - <f>)^2.
\]
The cross-correlation coefficients between two parallel spectra can be
defined in the same way.  

Fig. 15 plots the mean coefficients vs. $l$ (in units of \AA ) 
of 1000 synthetic samples simulated in the redshift box
2.2 to 2.6 in the same model as in Fig. 12.  The graph in the insert
is the same plot but at very small separations. The LCDM model differs from
the CDM model in this figure only at small separation, so is shown as
the dashed curved in the insert graph.
The 1$\sigma$ error bars in the CDM model are plotted as 
the two dotted curves. The heavy solid curve is from the Keck observation
of HS 1700+64 in the same redshift range.

Significant correlations are found within 10\AA 
\ (or 3.94 $h^{-1}$Mpc in comoving space in the CDM model), which should be 
basically attributed to individual Voigt profiles and density wings.
The data and simulations match each other within the $2\sigma$ error
bars.
If the wing does not exist, i.e., clouds are all zero-sized, the
coefficients will drop to zero at a smaller separation
(see Fig. 8 of Bi 1993 and Fig. 13 of MCOR).
Although the initial density distribution contains fluctuation powers
on scales
$> 10 h^{-1}$Mpc, no correlation can be found in the spectral 
coefficients on these large scales because the signals are too weak. 

\subsection{Spectral Filling Factor}

A useful statistical quantity is the cumulative distribution of the 
transmitted flux (see MCOR), or {\em spectral filling factor}. The 
observational calculation of the Keck spectrum is shown as the heavy 
solid curve in Fig. 16. The light solid curve and the dotted curve are    
from synthetic spectra in the CDM and LCDM models, respectively.
In order to mimic the observation best, the simulation samples were   
performed in the redshift interval 2.2 to 2.6;
in each pixel independent Gaussian noise of zero mean and 
variance 0.04 has been added to match the noise in the Keck spectrum. 
The statistical 1$\sigma$ error bars are 
so small that we may ignore them in the figure.
Noise in the simulation gives a little bit more filling factor at large
flux and a little bit less filling factor at small flux. 

The filling factor of the CDM model agrees well with that observed except 
for a small difference $\sim 0.027$ at large fluxes. As we have pointed 
out early in the paper, the present IGM model contains very few or no 
Lyman limit systems or damped systems. But the real Keck spectrum
does. Since these clouds have always absorbed flux to zero, adding them 
to the model will account for the difference to allowed error bars.
The LCDM model has stronger clustering and thus slightly less low 
density clumps, or equivalently, less shallow spectral absorptions. 
The density fluctuation in this model is 1.25, compared to 1.18 in the 
CDM model. Both the models are successful in explaining the existence
of very weak absorption features.

In an IGM with a very high variance, most of the
mass would reside in discrete 
clumps (or clouds) and most of the volume would be nearly 
empty
as underdense regions are destroyed. Due also to the 
saturation of overdense clouds in the Ly$\alpha$ absorption,
such a medium would require a lower UV background radiation
to give the correct HI Gunn-Peterson opacity. (For the HeII GP opacity,
the medium would require a very high population ratio.) This is the case
for the old, ``standard'' cold-dark-matter (SCDM) model 
with $\Gamma = 0.5$, which has $\sigma = 2.52$
at $z=2.4$. In order to get the correct effective opacity, 
we need to use $J_{21} = 0.005$,
which is much too small to be consistent with the proximity effect.
The strong clustering of the model will also produce a very
different spectral filling factor function, as shown in Fig. 16.

\begin{figure}[h]
\vbox to3.0in{\rule{0pt}{3.0in}}
\includegraphics{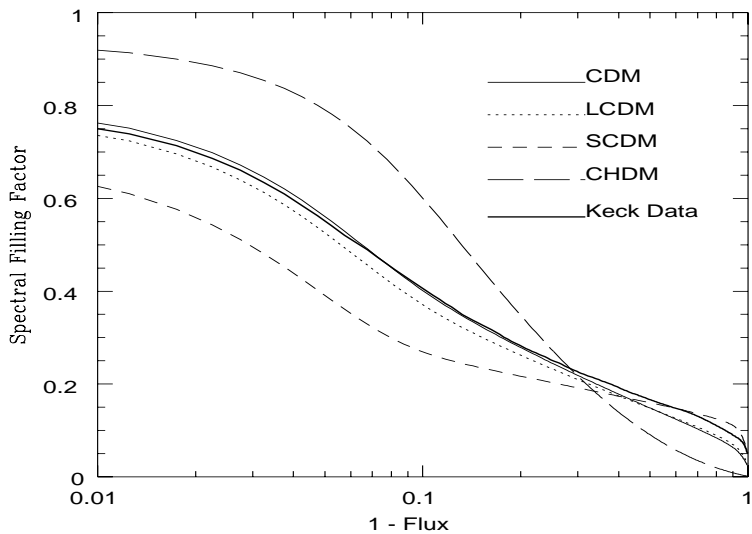}
\caption{Spectral filling factor vs.
the flux decrement in the CDM (solid line), LCDM (dot line),
SCDM (dashed line) and CHDM (long dashed line) models. The heavy solid
curve is from the Keck observation of HS 1700+64. The theoretical curves
have all had noise added to approximate the noise in the data for
HS 1700+64. \label{fig16}}
\end{figure}

On the other hand, an IGM with too low a variance would have difficulty in
making line-like features in the Ly$\alpha$ absorption because of 
insufficient density variations within clouds. Since more clouds are
not saturated, we need a higher UV flux for the correct
ionization and the effective opacity. However, the absorption would become
severely blended even at low redshifts. For example, the
cold-plus-hot dark-matter (CHDM) model with $\Omega _{hot} = 0.3, 
\Omega _{cold}  = 0.7$ (e.g. Klypin et al. 1993) has
$\sigma = 0.64$ at $z=2$. We need $J_{21} = 0.16$ to get the
correct effective opacity, but the absorption spectrum will
look too blended to be acceptable in the simulations. We expect to see
a different spectral filling factor function too.

The spectral filling factors in the SCDM and CHDM models are plotted
as the dashed and long dashed curves in Fig. 16. Although we can change
$J_{21}$ to obtain the observed opacity, these model distributions in the 
figure are completely inconsistent with the observation.
We thus conclude that the
$\Gamma = 0.3$ CDM model and the 
LCDM model are favored not only by other
cosmological considerations but also by
Ly$\alpha$ forest observations. The $\Gamma = 0.5$ SCDM
model and the CHDM model fail in the test of the
Ly$\alpha$ forest.

\subsection{Comparisons with MCOR}

Besides using real observations, we can compare our LCDM model with the same
model in MCOR in order to check how it 
compares with hydrodynamical simulations. 
Some of the comparisons have already been discussed above (e.g Fig. 4c and 
Fig. 8). The spectral filling factor function provides another excellent
comparison.  Fig. 17 shows such filling
factors at three redshifts $z=2$, 3 and 4 in our LCDM 
model. Unlike Fig. 16, these curves have no added noise.
Compared to MCOR's Fig. 11b, our curves predict slightly
more pixels with fluxes close to 1 because of somewhat
stronger mass clustering in our simulation. 
 
\begin{figure}[h]
\vbox to2.8in{\rule{0pt}{2.8in}}
\includegraphics{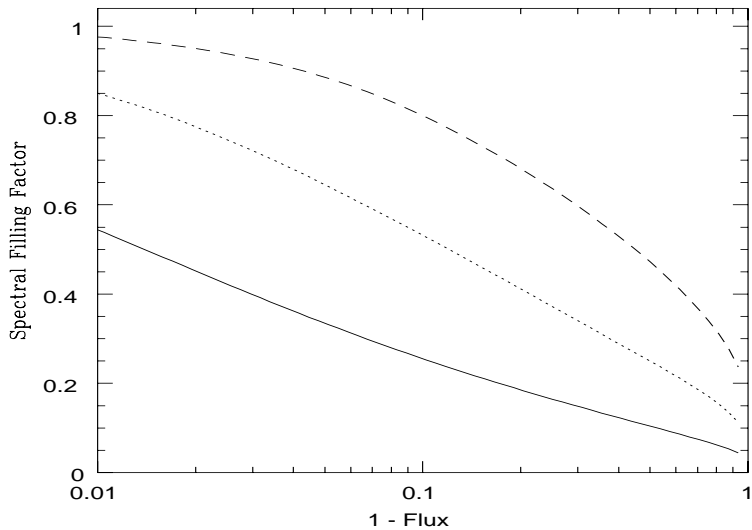}
\caption{Spectral filling factor vs.
the flux decrement
in the LCDM model. The solid,
dotted and dashed curves are for $z=2$, 3 and 4, respectively.
These curves may be compared with similar ones given by MCOR's
Figs. 11b. \label{fig17}}
\end{figure}

Fig. 18 is a plot of the mass fraction coming from pixels that have
transmitted flux smaller than a threshold value $F$ (see Fig. 21 of
MCOR). MCOR have a different procedure to simulate a spectrum, i.e.
they first transform the density distribution into velocity space 
using the peculiar velocity and then calculate the absorption spectrum 
regardless of the peculiar velocity. Therefore, the mass fraction they
derived in their Fig. 21 is in velocity space, instead of the
original comoving space.
In order to account for the difference, we use zero velocity in the 
calculation of Fig. 18. 

Our plots agree with the corresponding plots of MCOR
in general, although we are using a quite different approach to the
problem. There are two other factors that may bring uncertainties
into the comparisons.
First, MCOR did not use the Voigt profile,
but used the pure thermal Gaussian profile
in their spectral calculation.
Second, the grid size in their simulation is 34$h^{-1}$kpc in comoving
space, so their spectrum cannot be resolved below this limit
(4.4$h^{-1}$ km s$^{-1}$ in velocity). Both of these
factors can account for
the small differences between the results of our calculations and theirs.

\begin{figure}[h]
\vbox to2.8in{\rule{0pt}{2.8in}}
\includegraphics{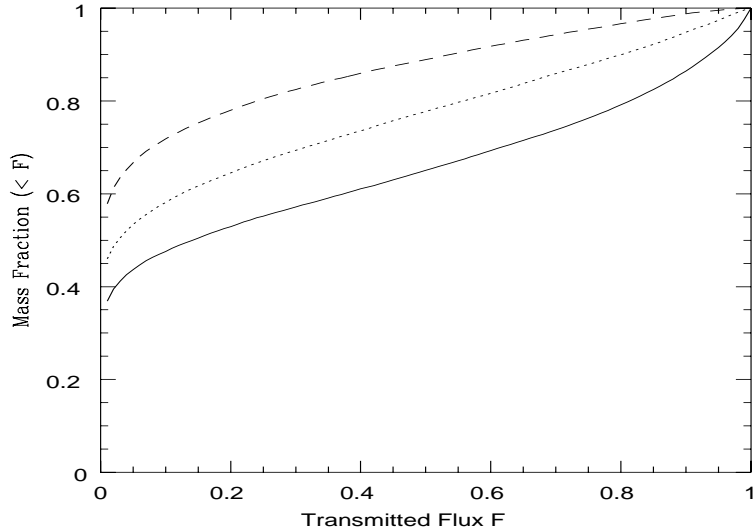}
\caption{Mass coming from pixels with transmitted flux
less than a given threshold at $z=2$ (solid), 3 (dotted) and 4 (dashed)
in the LCDM model.
This figure may be compared with a similar one given by MCOR's
Fig. 21. \label{fig18}}
\end{figure}

\section{Summary}

In this paper, we have argued that the intergalactic medium should be
regarded as a continuous medium with density fluctuations that result
from the growth of primordial dark matter fluctuations, as opposed to
the conventional view of discrete clouds immersed in a smooth intercloud
medium. With this paradigm shift, the long-held distinction between the
Gunn-Peterson effect (arising from the diffuse intercloud medium) and
the Bahcall-Salpeter effect or Ly$\alpha$ forest of discrete absorption
lines (arising from denser clouds) no longer makes sense. The dividing
line between these aspects of the IGM is arbitrary. If one were to use
the universal mean baryon density to distinguish between ``clouds'' and
the ``intercloud medium'', for example, this would correspond to absorption
features with an estimated column density of about $10^{13.5}$ cm$^{-2}$
at $z=3$. Lower column density features typically come from regions that
are below the mean density. At $z=3$ our model indicates that about
30 percent of the mass, occupying 70 percent of space, would fall in this
category. Most of the baryonic mass is in the 
moderately overdense fluctuations that 
produce the well-studied Ly$\alpha$ forest.

We summarize our main conclusions further as follows.

1) The nature of the IGM at $z=2$ to $z=4$.

From both our analytical calculation and the hydrodynamic simulation
of Cen et al. (1994) and MCOR, we conclude that the fluctuation variance 
of the density field of the IGM is close to unity in the redshift
range ($z=2\to 4$) where most of the Ly$\alpha$ forest is observed.
This is a special phase in the evolution of the
IGM when both overdense and
underdense clumps are significant in appearance. Mass is moving 
generally from low densities to high densities, but has not yet arrived
at the final virialized stage when the clumps are stable in proper
space. Dark matter fluctuations would act like magnets that attract
the baryonic medium to their potential wells, but do not necessarily
confine the baryons. Pressure confinement cannot be achieved.
There are no distinguished thermal phases in the medium from which 
we can discriminate different thermal components. The whole medium 
appears as one diffuse and continuous fluid. 
This median fluctuation of the IGM is important in producing the correct
HI opacity and the number of Ly$\alpha$ lines observed.
The lognormal hypothesis appears to provide an excellent description
of the density distribution of intergalactic gas in the non-linear
regime at $z<4$.

2) The structure within Ly$\alpha$ clouds.

In the diffuse IGM, our term ``clouds'' actually refers to 
all fluctuated density clumps which have either overdensity above
the mean or underdensity below the mean. 
In fact, typical clouds and thus typical Ly$\alpha$ lines have 
just the mean mass density. For example, at redshift 3, clouds of
column density $10^{11.5}$, $10^{13.5}$ and $10^{15.3}$ cm$^{-2}$ have 
mass density that is 0.1, 1 and 10 times the mean. 
The variation of (neutral) density within a cloud is best 
described by an exponential profile that peaks at the cloud center 
and declines outside the Jeans length. The origin of the 
density wing is actually due to the fluctuation nature of the IGM in
which we cannot distinguish one cloud from another in space.
However, the density wing has a very important implication : it broadens
the absorption profile significantly. A wing of any absorber will 
resemble another cloud with a lower column density. For instance,
a cloud with $N_{HI} = 10^{14.5}$ cm$^{-2}$ has the same
neutral hydrogen distribution as an $10^{12.5}$ cm$^{-2}$ cloud
at the comoving wing distance 0.8 $h^{-1}$Mpc, so it would have the detectable
size 0.8 $h^{-1}$Mpc.

The peculiar velocity causes mass to fall toward the cloud center where
the dark matter potential is minimized in comoving space. However,
all clouds below $10^{16}$ cm$^{-2}$ are still expanding in proper
space, because the peculiar velocity is not large enough to overcome the
Hubble velocity. For clouds above $10^{13.5}$ cm$^{-2}$, the velocity 
is comparable to the thermal $b$ parameter, so the resulting profile
will be narrower. This effect is just opposite to the density wing.
For an individual cloud, the velocity could appear as turbulent motions
because it is not necessarily symmetric.

One effect that has been neglected in the present study is shock waves in 
the central regions of high column density absorbers, which can 
be anticipated in the simple Zel'dovich approximation and have actually 
been observed in hydrodynamical simulations (e.g. Cen et al. 1994). 
Two mass streams from opposite directions along the principal axis
are falling toward the center of a pancake. If thermal pressure cannot 
overcome the motions, they will impact each other and leave a cooling mass 
layer at the cloud center. Such a cooling layer will produce extra neutral
hydrogen and thus will enhance the Ly$\alpha$ absorption opacity. 
However, a statistical histogram of
density vs. temperature in the simulation of MCOR showed this is 
important only
for clouds with $N_{HI} \ge 10^{15}$ cm$^{-2}$. Most of Ly$\alpha$
forest cloud discussed in this paper will not be affected by 
such shock waves.

3) The number density of clouds.

The total number of all clumps in the IGM is determined by the number
of dark matter potential wells when smoothed on the Jeans scale. Since
the Jeans length is roughly a constant in comoving space, the ``global''
number will be more or less fixed. The same cloud with a high column
density  will evolve to lower and lower column densities at later times
because the proper density is decreasing. The column density distribution
function at one redshift can be identified with that at a lower redshift
by shifting the column density by a constant decreasing factor.
Thus we expect to have more low column density
absorbers at $z=2$ than at $z=4$. 

Our simulation is based on the assumption that the UV background intensity
is constant between $z=4$ and $z=2$, yet the effective opacity
and the number counts of Ly$\alpha$ lines 
are quite consistent with those observed.
Therefore
we do not need to introduce any evolution of the UV 
background in that range.

4) The baryonic density

For both the CDM and LCDM models considered here, matching the observed
effective opacity in the Ly$\alpha$ forest requires an appropriate choice
of $\Omega _b ^2 / J_{21}$. If we assume $\Omega _b = 0.015h^{-2}$, a
nominal value consistent with standard Big Bang nucleosynthesis, then
both models require $J_{21} \simeq 0.2$, somewhat lower than the value 
estimated for quasars at $z=2 - 3$, and substantially lower than the
estimate that $J_{21} = 1.0 ^{+0.5} _{-0.3}$ from the proximity
effect (Cooke et al. 1996). Alternatively, if we adopt $J_{21}
= 0.5$ as estimated by Haardt \& Madau (1996), then we require 
$\Omega _b = 0.025h^{-2}$. This is at the high end of the range of 
values consistent with standard Big Bang nucleosynthesis. It is, however,
compatible with the value found recently by Tytler, Fan \& Burles (1996)
from a measurement of a low deuterium abundance at high redshift. Our
results are obviously incompatible with the much lower value of 
$\Omega _b$ deduced by Rugers \& Hogan (1996) from a measurement of 
a high deuterium abundance.

5) Statistics of absorption spectra

We have compared the Ly-$\alpha$ 
forest of HS 1700+64 with synthetic samples in the CDM and LCDM models 
using several statistical quantities which depend sensitively on spectral 
shapes. The $b$ parameter describes the temperature or the width of an
individual line profile. The spectral filling factor represents the 
percentage of pixels with flux below a given limit. 
It is a useful quantity that can be measured in observations. 
A strong clustering in the
IGM would result in too small a filling factor to be accepted.
The spectral autocorrelation coefficient reveals the 
correlation between flux at chosen separations. The most useful 
coefficients are within 10\AA \ and are dominated by
the Voigt profile and density wing. 

6) Related studies.

While our model could still be modified somewhat in detail, the
global picture and physical explanation of the IGM and the Ly$\alpha$ 
forest should be well described in the 
present calculation. There are several interesting subjects 
in the IGM models.

First, highly resolved Keck spectra would allow us to study small
scale structures in the IGM, and to discriminate different cosmological
parameters. Beside the traditional Voigt fitting, statistics like the 
spectral filling factor and the autocorrelation function would
be easy to compute and even more clear in revealing other important
absorption features, especially weak features. The Ly$\alpha$ forest
is very useful in model diagnostic checking. For example, if the 
standard deviation $\sigma$ of the IGM
is changed by a factor of 2, such as in the 
$\Gamma = 0.5$ CDM model and the cold-plus-hot 
dark matter model, the resulting
absorption spectra are already inconsistent with observations and the
models can probably be ruled out.

Second, the present IGM model is able to explain the recent observations
of the HeII Gunn-Peterson effect with a reasonable HeII to HI population
ratio of 100 - 150. We will discuss this in detail in another paper 
(Davidsen et al. 1996). 
The HeII absorption depends not only on the column density of a 
cloud but also on the cloud geometry because of different saturation
in the curve-of-growth. 
The more extended the cloud is in space, the less saturation the
absorption has.
In the conventional Ly$\alpha$ cloud picture, a cloud with $10^{13}$ 
cm$^{-2}$ column density is already saturated in the HeII absorption.
In the present model, because of extended spatial 
structures, most of such clouds still appear linear 
and thus make a dominant contribution in the HeII opacity.

Finally, we note that if the Ly$\alpha$ forest originates from
continuous density variation
in the IGM, one can in principle deconvolve the density
field inversely from the forest. This will allow us to obtain very
fine structures ($\sim 0.05h^{-1}$Mpc) as well as large scale 
structures of the cosmic baryonic distribution at high redshifts for
the first time.

\acknowledgments

We thank J.R. Bond, B. Espey, G. Kriss, J. Miralda-Escud\'e, C. Norman, 
J. Peebles, Y. Pei, J. Ostriker, D. Weinberg 
and W. Zheng for many stimulating discussions and suggestions. HGB is
especially grateful to W. Zheng for his help in using the Voigt fitting
software ``VPFIT''. We thank D. Tytler for generously providing
the Keck spectrum of HS1700+64. This work has been supported by NASA
contract NAS 5-27000 to the Johns Hopkins University.

\clearpage

\begin{table*}
\begin{center}

\begin{tabular}{|l|l|l|l|l|l|l|} \hline\hline
$N_{HI}$    & \multicolumn{3}{c|}{$\log{M/M_{\odot}}$, \em CDM} &
              \multicolumn{3}{c|}{$\log{M/M_{\odot}}$, \em LCDM}\\
\hline
                   & $z=2$  & $z=3$ & $z=4$ & $z=2$ & $z=3$ & $z=4$  \\
\hline
$10^{16}$ cm$^{-2}$ & 12.2  & 11.6  & 11.2  & 12.3  & 11.8  & 11.4  \\
$10^{15}$ cm$^{-2}$ & 11.3  & 10.9  & 10.5  & 11.5  & 11.0  & 10.7  \\
$10^{14}$ cm$^{-2}$ & 10.7  & 10.2   & 9.8   & 10.8  & 10.4   & 9.9   \\
$10^{13}$ cm$^{-2}$ & 10.0  & 9.5   & 8.9   & 10.2  & 9.6   & 9.1   \\
$10^{12}$ cm$^{-2}$ & 9.3   & 8.6   & 8.1   & 9.5   & 8.8   & 8.3 \\
Jeans Mass          & 10.2  & 9.9   & 9.7   & 10.5  & 10.2  & 9.9 \\
\hline \hline
\end{tabular}

\end{center}
\caption{Baryonic Mass Associated with Typical Features in 
the Ly$\alpha$ Forest \label{tbl-1}}
\end{table*}

\clearpage

\ \ 
\begin{table}
\dummytable\label{tbl-2}
\end{table}

\clearpage
 
\begin{table*}
\begin{center} 
\begin{tabular}{|l|l|l|l|l|l|l|l|l|l|} \hline\hline
Line & \multicolumn{3}{c|}{$z=4$} & \multicolumn{3}{c|}{$z=3$} &
       \multicolumn{3}{c|}{$z=2$} \\
\cline{2-10}
     & $\log N_{int}$ & $\log N_{fit}$ & $b_{fit}$ 
     & $\log N_{int}$ & $\log N_{fit}$ & $b_{fit}$
     & $\log N_{int}$ & $\log N_{fit}$ & $b_{fit}$ \\
\hline
1    & 14.05 & 14.16 & 62.6 & 13.39 & 13.54 & 55.0 
     & \multicolumn{3}{c|}{see\tablenotemark{3}} \\
2    & 14.83 & \multicolumn{2}{c|}{see\tablenotemark{1}} 
     & {see\tablenotemark{2}}& 13.69 & 272.0
     & {see\tablenotemark{2}}& 13.44 & 77.8 \\
3    & 15.14 & 15.49 & 55.5 & 14.91 & 15.07 & 46.5 & 14.34 & 14.35 & 36.4 \\
4    & 13.57 & 13.67 & 38.6 & 12.92 & 12.94 & 30.0 
     & \multicolumn{3}{c|}{{see\tablenotemark{3}}} \\
5    & 14.35 & 14.31 & 26.1 & 13.85 & 13.59 & 23.0 & 13.17 & 12.76 & 19.1 \\
6    & 14.33 & 14.51 & 77.3 & 13.76 & 13.99 & 55.6 & 12.92 & 13.30 & 44.5 \\
\hline  
\hline
\end{tabular}
\end{center}

\tablenotetext{1}{Because of blending and saturation, the line cannot be 
found by
Voigt fitting in the spectrum.}

\tablenotetext{2}{Selected only in the ``no smooth'' identification. 
The blended asymmetric absorption profile requires line 2 in the Voigt 
fitting, which, however, is not cloud 2.}

\tablenotetext{3}{The intrinsic absorber is merged into absorber 3.
         In the Voigt fitting, the line is not required either.}
\caption{Parameters of specific Ly$\alpha$ lines
at different redshifts
(intrinsic absorbers are selected according to the
``maximal smooth'' identification) \label{tbl-3}}

\end{table*}


\end{document}